\documentclass[manuscript]{acmart}
\AtBeginDocument{%
  \providecommand\BibTeX{{%
    \normalfont B\kern-0.5em{\scshape i\kern-0.25em b}\kern-0.8em\TeX}}}

\usepackage{footnote}

\usepackage{xcolor}

\begin{document}

\newcommand{\think}{Thinking Assistant }
\newcommand{\thinks}{Thinking Assistants}

\title[\emph{\thinks{}}: LLM-Based Conversational Assistants that Help Users Think By Asking]{\emph{\thinks{}}: LLM-Based Conversational Assistants that Help Users Think By Asking rather than Answering}

\newcommand{\SP}[1]{\textcolor{purple}{\textbf{*Soya*}: #1}}

\author{Soya Park}
\email{soya.park@emory.edu}
\affiliation{%
  \institution{Emory University}
  \country{USA}
}

\author{Hari Subramonyam}
\email{harihars@stanford.edu}
\affiliation{%
  \institution{Stanford University}
  \country{USA}
}

\author{Chinmay Kulkarni}
\email{chinmay.kulkarni@emory.edu}
\affiliation{%
  \institution{Emory University}
  \country{USA}
}

\renewcommand{\shortauthors}{Soya Park, Hari Subramonyam and Chinmay Kulkarni}

\begin{abstract}

Many AI systems focus solely on providing solutions or explaining outcomes. However, complex tasks like research and strategic thinking often benefit from a more comprehensive approach to augmenting the thinking process rather than passively getting information.
% In an increasingly automated world, AI is often seen as providing answers or completing tasks. However, many tasks, such as research and strategic thinking, benefit more from guidance and critical inquiry than direct solutions.
We introduce the concept of ``\thinks{}'', 
% an approach that engages in deep reflection and critical thinking through thought-provoking queries. Specifically, we introduce 
a new genre of assistants that help users improve decision-making with a combination of asking reflection questions based on expert knowledge. Through our lab study (N=80), these Large Language Model (LLM) based \thinks{} were better able to guide users to make important decisions, compared with conversational agents that only asked questions, provided advice, or neither.  

Based on the results, we develop a \think in academic career development, determining research trajectory or developing one's unique research identity, which requires deliberation, reflection and experts' advice accordingly. In a longitudinal deployment with 223 conversations, participants responded positively to approximately 65\% of the responses. 

Our work proposes directions for developing more effective LLM agents. Rather than adhering to the prevailing authoritative approach of generating definitive answers, LLM agents aimed at assisting with cognitive enhancement should prioritize fostering reflection. They should initially provide responses designed to prompt thoughtful consideration through inquiring, followed by offering advice only after gaining a deeper understanding of the user's context and needs.

%In such situations, juniors often lack direct mentorship from professors, or may feel hesitant to approach faculty with their queries, making \think{} particularly useful. We thus present \Sys{}.
% Leveraging a Large Language Model (LLM), \Sys{} is a demonstration system built as a \think{} for working with specific professors in the field of human-computer interaction (HCI). It was designed through training on information specific to these professors and validation processes in collaboration with these academics. This paper delineates the system's architecture and offers an analysis of our deployment study. Additionally, this paper covers the spectrum of questions posed to our chatbots by users. The system recorded 223 conversations, with participants responding positively to approximately 65\% of responses. We also conduct exit interviews with seven users. Our findings indicate that users who discuss and brainstorm their research interests with \Sys{} engage more deeply, often interacting with the chatbot twice as long compared to those who only pose questions about professors. We also find that participants who use the system as an informational or predictive agent, seeking definitive answers such as a direct 'yes' or 'no' to queries like \textit{``Do I have a chance of getting into the university?''} find the \think's reflective conversations unsatisfactory. 
  
\end{abstract}

\newcommand{\Sys}{\textsc{Gradschool.chat}}

% \includegraphics[width=\linewidth]{figs/workflow2.pdf}
% \caption{\Sys{} is a \think that helps prospective graduate students reflect on research interests. Based on the user's question, the assistant chooses between a `probing' mode, understanding user interests and prompting reflection, and an `answering' mode that offers relevant information. Most conversational turns end with an encouraging follow-up question.}
% \label{fig:workflow}
  
\maketitle

\section{Introduction}

% Helping users' critical thinking is a multifaceted endeavor that intersects with various objectives of human-computer interaction. A key opportunity is to scaffold reflection. While extremely valuable for x, y, z[cite], reflection is difficult, even for seasoned experts or eminent figures in their respective domains~\cite{cohen2010take}. Then, when a person confronts substantial decisions such as transitioning jobs or evaluating multiple job offers, they encounter the arduous task of conducting comprehensive comparisons among the available options. Finally, if the user is a newcomer, the lack of guidance exacerbates the difficulty of elaborating on and deliberating over their choices. Good mentors or life coaches aid these tasks; good artificial intelligence helps as well, but it is clear that we have far to go. 
Individuals often struggle with reflective thinking, especially when facing substantial decisions such as career transitions or evaluating competing job offers. This difficulty stems in part from the complexity of conducting comparisons along multiple dimensions, only some of which are readily apparent. Such decisions are frequently associated with rumination, i.e. conscious, repetitive thought, but rumination does not always lead to better or more creative decision-making~\cite{cohen2010take,treynor2003rumination}. Critically, for better decision-making, individuals need to engage in productive self-reflection. i.e. repeated thinking that allows for elaboration of ideas, and one's relationship to them~\cite{verhaeghen2005whywesing}.  For non-experts and newcomers (such as graduate students considering alternative research careers, or those transitioning careers), this is especially hard, as their lack of expert knowledge prevents elaboration and deliberation, and their outsider-status leads to fewer social connections that can scaffold well-informed self-reflection.  

This paper introduces a genre of AI assistants, \textit{\thinks{}}, that scaffold productive self-reflection.  \thinks{} are inspired by two key observations: first, 
% 
% , a key aspect of critical thinking that intersects with various objectives of human-computer interaction. While traditionally good mentors or life coaches have aided in these tasks, artificial intelligence presents a promising avenue for support. By leveraging AI to guide users through structured reflection processes, we can potentially enhance decision-making capabilities across various levels of expertise. This approach shows particular promise in [specific areas x, y, z - citation needed], although significant advancements are still needed to fully realize AI's potential in supporting critical thinking and decision-making processes.
% 
in our everyday life, when we need help with complex cognition and manipulation tasks, we often turn to other people. We ask experts or colleagues (i.e. experts who know well about us) to think through together that we cannot answer ourselves -- whether we hope to navigate careers or buy a car~\cite{schotter2003decision}, or improve our mental health ~\cite{davidson2005recovery}, we rely on others,  not to make decisions for us but how to explore ourselves to reach the best decisions.  By talking to individuals who are knowledgeable about the subject of concern, one can explore different perspectives and restructure their thoughts through articulation~\cite{aakhus2007conversation, schon1984reflective}. A second unfortunate observation is that engaging in a thorough discussion about one's personal decision is often costly or inaccessible, accessible only to those with a robust network, not to those in a peripheral position (i.e., newcomers or emerging professionals)~\cite{financeStrategists2023, rothschild2022datawork, aakhus2007conversation}. %~\cite{weinstein2018teaching, higgins1978saying}. -> edu citation, largely about knowledge acquisition

\thinks{} build on both these observations. The key insight they leverage is that recent large-language models can engage in dialogue that encourages reflection, and also encode nuanced information about the world (so-called ``world-knowledge''), but can be augmented with specific expertise. Specifically, we hypothesize that Large Language Model (LLM) agents with a basic knowledge of a domain, augmented by expert human knowledge, can support users' reflective thinking through reflective inquiry. Through interactive discussions, these assistants engage users in exploring and deepening their understanding of various domains, facilitating a comprehensive exploration of the subject matter at hand.

\think is comprised of two main components:
\begin{enumerate}
    \item Reflective inquiry: The primary objective is to probe users' interests, emphasizing active engagement over passive reception of information; to promote reflective thinking through  inquiry. 
    \item Expert knowledge: Inquiry alone is insufficient if it lacks grounding; effective inquiry requires more than merely posing questions for articulation~\cite{aakhus2007conversation}. Thus, \think employs a framework where inquiry is  constructed and guided by domain-specific expertise.
    % \item Wise feedback assistant: Once the agent asks enough questions and gets a good sense of the user, it provides \textit{wise feedback} to aid self-reflection~\cite{walton2018wise}, i.e. offering constructive feedback based on high standards and affirming the users' capabilities to meet these standards, and recommends relevant research papers. 
    % \item Safety bot: To prevent factual inaccuracies, we also additionally implemented safety measures, including a secondary virtual assistant that corrects any discrepancies in the chat messages. 
\end{enumerate}

This paper makes two contributions: first, it demonstrates that for scaffolding productive self-reflection, \thinks{} must be designed with an \emph{emphasis on inquiry}. This is in contrast to the traditional emphasis in the design of such agents; one of furnishing answers, either through their encoded knowledge, or through retrieving information from  knowledge bases. A \think \emph{poses thought-provoking questions and cultivates a dynamic exchange that stimulates user interest and encourages active participation}. Asking such questions requires domain knowledge, but their goal is to help users understand, refine, and develop their own interests, rather than to ``transfer knowledge'' (in this sense, \thinks are constructivist learning aids~\cite{cowan1969social}.) Second,  we demonstrate the effectiveness of \thinks. Through a controlled experiment (N=80) to show that participants engage in reflective thinking more frequently while using a \think than using assistants that either only provide expert advice or ask reflective questions not informed by domain expertise. Further, we evaluate a particular \think in a longitudinal deployment. This assistant, \Sys{} is designed to support students and junior researchers in choosing research directions. In this longitudinal study, users had 223 anonymous conversations about their graduate school choices. Most users naturally end up resorting to traditional conversational interactions with agents, expecting direct answers due to the prevalence of this interaction style in most existing agents. This expectation is shaped by the dominant conversation paradigm. However, users who remain aligned with this conventional approach, rather than embracing informed inquiry, often find the interaction less helpful. They tend to exhibit significantly lower levels of engagement -- approximately half as much compared to those who adapt to or explore a more inquiry-driven conversational flow, which encourages deeper understanding and active participation.

\section{Background \& Related Work}

The scope of our work touches upon several aspects of learning at scale and conversational assistants that have been explored by prior researchers in three main areas. We first describe the previous approach to self-discovery and guidance on reflective thinking. We then review the methods of ensuring the safety of conversational agents, which are our means of scaling the thinking process especially relevant as we are focusing on reflective environments (e.g., graduate-application process). Lastly, we focus on previous work relevant to \Sys{} on scaling mentorship and guidance to students, especially in the case of lacking teachers and mentors.

\subsection{Systems for Self-Discovery}

In open-ended domains where definitive answers are elusive, decisions are often guided by self-reflection and self-discovery (\textit{Who am I? What do I like?}) by affirming their values and choices~\cite{sherman2006psychology}. Self-discovery is developing ``a self-system that explains ourselves to ourselves, a system that persistently explains our behavior, and the world at large, so as to sustain a phenomenal experience of the self as adaptively and morally adequate''~\cite{Steele1988selfaffirmation}. These processes empower individuals to navigate ambiguity and chart their own course based on personal values, aspirations, and insights~\cite{waterman1984identity, Hinkle2020SelfExploration}. 

Self-reflection poses a significant challenge as it requires a deep analysis and examination of oneself~\cite{berzonsky2011social}. Often, our own perspective may not fully illuminate aspects of ourselves that are apparent to others or essential for personal growth~\cite{park2023retrospector}. This discrepancy can hinder our ability to gain a comprehensive understanding of our strengths, weaknesses, and areas for improvement~\cite{Robb1967SelfDiscoveryAT}.

Instead of tackling self-reflection in isolation, individuals frequently seek assistance from others to identify their strengths and weaknesses. Collaborative reflection allows individuals to gain valuable external viewpoints, which can shed light on aspects of themselves that they may overlook or undervalue~\cite{aakhus2007conversation}. They often rely on colleagues for this purpose, drawing on their insights and perspectives~\cite{park2024tasklight, evans2008towards}. Alternatively, individuals may reach out to professionals, such as life coaches, who specialize in guiding individuals through the process of self-discovery and personal growth~\cite{Shaw2018TheEO}. They help individuals identify their core values, passions, and aspirations. They facilitate deep introspection to clarify what truly matters to the individual and what they want to achieve in various aspects of their life. 

HCI communities have investigated various systems to support self-reflection through different interventions. These approaches include provoking thoughts about oneself by introducing different personas and having conversations with persona-driven chatbots~\cite{lee2020hear}, encouraging users to think from different perspectives~\cite{yeo2024help}, collecting personal information to gain insights~\cite{xu2024understanding}, and sharing this information with others~\cite{wang2022snappi}. LLM-powered chatbots have emerged as a promising avenue for self-discovery, equipped with capabilities to provide interactive support, such as talking human-like (sharing experience or opinion on topics) and remembering essential user information~\cite{lee2020hear, jo2024LTM}. While authentic interactions with chatbots allow users to feel more connected, they also increase the risk of over-reliance and frustration due to the inherent gap between human communication and chatbot capabilities~\cite{ma2023chatbot, jo2024LTM}, as chatbots can not fully grasp the nuances of human interactions~\cite{ma2024lgbtq}. Instead of focusing on imitating the tone of experts, our approach leverages the power of LLMs to perform reflective inquiry, aiding in self-discovery and promoting reflective thinking through questions that experts might ask. 

\subsubsection{Asking rather than Answering}

Asking questions has long been a popular and effective method in educational settings, facilitating active learning and helping students elaborate and clarify their knowledge~\cite{weinstein2018teaching, higgins1978saying, dewey2022we}. By responding to questions posed by teachers, students engage more deeply with the material and develop a better understanding. Recent research indicates that answering questions generated by not just teacher, but LLMs can also foster robust logical thinking~\cite{danry2023ask}. The effect of questions extends to self-discovery and personal growth as well. Engaging in reflective inquiry allows individuals to ``conduct a crucial experiment in the form of a question''~\cite{aakhus2007conversation}. Since life decisions are often open-ended and filled with tacit information, there are no definitive right or wrong answers. Instead, the process involves considering multiple questions and perspectives~\cite{schon1984reflective}. 

Asking contextual questions and personalizing advice is inherently resource-intensive because it requires a deep understanding of individual circumstances, preferences, and goals. This process often demands significant time, expertise, and computational power, especially when scaled across large populations~\cite{bhatt2024can}. Existing AI-assisted decision-making tools primarily focus on providing information, rather than fostering deeper thinking and reflection~\cite{lai2021towards}. While these systems can efficiently deliver generic recommendations, understanding users' intents and adapting advice accordingly demands significantly more resources and effort~\cite{andukuri2024star}. This challenge shapes the mental models of users, who may begin to view AI as primarily transactional and informational, rather than as a tool for deeper cognitive engagement~\cite{Luger2016CA, zaib2022conversational}. 

Recent advancements in technology, such as LLMs, have demonstrated a promising ability to ask meaningful questions and gain a deep understanding of users. Leveraging these technologies, survey systems can interactively engage with users to elicit more thoughtful and accurate responses~\cite{xiao2020tell}, while also capturing implicit signals such as users' preferences~\cite{andukuri2024star}. Beyond merely gathering information, this process has been shown to benefit users by promoting logical thinking~\cite{danry2023ask}. However, there remains limited research on how question-asking systems designed for open-ended domains can support self-discovery and reflection encouraging deeper engagement and fostering self-awareness.

Now there is an opportunity to enable scalable interactions that combine the strengths of informed inquiry with domain expertise. We propose leveraging a primary LLM conversation model focused on asking questions rather than solely providing answers. Our hypothesis is that LLM agents engaging in interaction models designed to ask expert-level questions can stimulate users' deeper reflective thinking, fostering a more meaningful and thoughtful engagement process.

\subsection{Using Agents to Personalize Guidance}

% remember information about the users to provide personalized guidance~\cite{jo2024LTM}

% In the context of academic procrastination, the potential of LLMs lies not in replacing traditional methods like setting reminders or automatic scheduling, but in their ability to contextualize and intelligently integrate a broad spectrum of procrastination management strategies~\cite{bhattacharjee2024understanding}; personalized guidance option: option to specify the tone (formal vs. informal), directedness (direct vs. indirect), and length of the LLM-generated message (50, 100, or 150 words). They could choose whether or not to include specific instructions for addressing their situation, as well as select specific prompts for generating the message; Future procrastination management tools may occasionally restrict users’ access to functionalities that produce direct step-by-step solutions and instead prompt them to independently seek out potential strategies

Prior work has leveraged machine-learning-enabled chatbots to create personalized interactions at scale, especially in situations where there is limited availability of experts. For instance, in healthcare, to mitigate the limited availability of medical professionals, researchers have developed chatbots capable of handling basic health inquiries~\cite{bickmore2018safety, xiao2023powering}. Often, these systems help triage questions, allowing experts to focus on the most challenging issues. 
At the same time, such dialogue agents can also help in ways that humans cannot. For instance, in providing mental health support, such agents allow some participants to be more disinhibited in their conversation~\cite{ma2023chatbot}. Thus, in addition to interpreting natural language queries and accessing knowledge bases,  the non-human nature of these agents may be helpful in applications where people may be hesitant in asking others for help~\cite{ma2023chatbot, sharma2023reframing}. 

In educational settings, most chatbots are designed to teach new skills, enhance learning efficiency, motivate learners, and scale educational offerings~\cite{wollny2021we}. However, the support provided by these chatbots has primarily focused on delivering information rather than facilitating students' reflective thinking or helping them find the right fit for their educational needs~\cite{bhattacharjee2024understanding, kasneci2023chatgpt}. For instance, while intelligent agents are adept at helping students navigate complex administrative logistics and academic requirements, they primarily serve as a substitute for limited administrative staff, focusing on information provision rather than advancing students' academic exploration~\cite{mitra2023managing, LefdahlDavis2018TheIO}.
\thinks{} share similarities with these existing educational systems in offering personalized guidance but distinguish themselves by focusing on reflective thinking through thought-provoking questions.

Our work is also motivated by prior work that finds that outsiders, such as junior researchers often find it stressful to interact with more knowledgeable community members, such as senior academics, leading to over-preparation for meetings~\cite{park2023who2chat}. By contrast, \thinks{} offer a more relaxed setting where users can be less inhibitied, and freely explore and ask questions without fear of embarrassment. 
% they might hesitate to pose to professors for fear of being seen as unprepared. Such exploration can then help students to determine their interest in specific research topics or graduate programs, and gauge their interest in reaching out to professors. Interacting with professors in turn offers unique benefits, such as gaining insights into research and building social capital. While not the focus of this study, we conjecture that \thinks{} may help students better prepare for such meetings in a less stressful way. 

\subsection{Conversational Agent Safety}
Ensuring the safety of conversational assistants, particularly those not designed for information seeking, remains a complex challenge. This is partly because the required safety measures vary depending on the assistant's purpose~\cite{Kempt2023AppropriatenessIA}. Most current measures focus on the accuracy of information provided by conversational assistants, which may not be entirely suitable for free-flowing conversations~\cite{zubatiy2023limitations}. 
% Our work diverges from the typical information-seeking focus. The primary objective of our system, \think, is to aid 
% users in their thinking process within the context of a specific experts' knowledge, as represented by their virtual assistant. 
In contrast, our focus is on asking well-informed thoughtful questions. Even so, it is often necessary to provide some reflective and frequently asked information for users. This is challenging because of the limitations of LLMs, specifically the potential for hallucination~\cite{dziri2022origin, ji2022hallucination}). 

Prior work has extensively explored various methods to improve factuality. One effective strategy involves structuring the conversation around expected domain-specific topics, and steering conversations away from other topics~\cite{Kempt2023AppropriatenessIA}. For instance, an agent designed for booking flight tickets might predominantly navigate branches related to gathering travel itinerary details, selecting tickets, and processing payments. This allows for more precise shaping and testing of the conversation flow for chatbot developers~\cite{klemmer2000suede, choi2021protochat}. Additionally, the incorporation of experts and inclusive datasets in training further bolsters the safety of conversational assistants~\cite{aroyo2023dices, xiao2023powering}.

While this prior work narrows the domains that the assistant deals with, we found that it is sometimes helpful to allow a \think to ask questions on broad areas such as the users' interest. Instead of narrowing the topics of conversation, we introduce a `safety bot' whose sole function is to verify the accuracy of the information provided by our virtual assistants in the narrow, targeted field of expertise. As such, our method allows the assistant to converse on a broad variety of topics, but only improves factuality in a narrow field.

\section{What kinds of assistant interactions lead to self-reflection?}

Informed by prior work, we hypothesize that users engage in reflective thinking most effectively when they interact with assistants that ask thought-provoking questions supplemented with domain experts' knowledge, in contrast to those that ask expertise-uninformed questions, or only offer expert advice. In our first lab study, we compared participants’ depth of reflective thinking in the presence of informed inquiry and/or advice. To allow quantitative comparison of the results of informed inquiry with or without expert guidance, we conducted subsequent dialog analysis with condition-blind crowd raters. 

\subsection{Experiment on different interaction modes}

\subsubsection{Participants}

We posted this task on Prolific (https://www.prolific.com/). We restricted our recruitment to adults who are fluent in English. They were paid \$5 per 20 minutes for their time and effort. We conducted a power analysis for a four-group ANCOVA study seeking a high power (0.8) with an alpha of 0.05 according to Cohen’s conventions, at medium to high effect size (0.38), yielding 20 participants per condition. 

\subsubsection{Procedure}

\begin{figure}
\includegraphics[width=\linewidth]{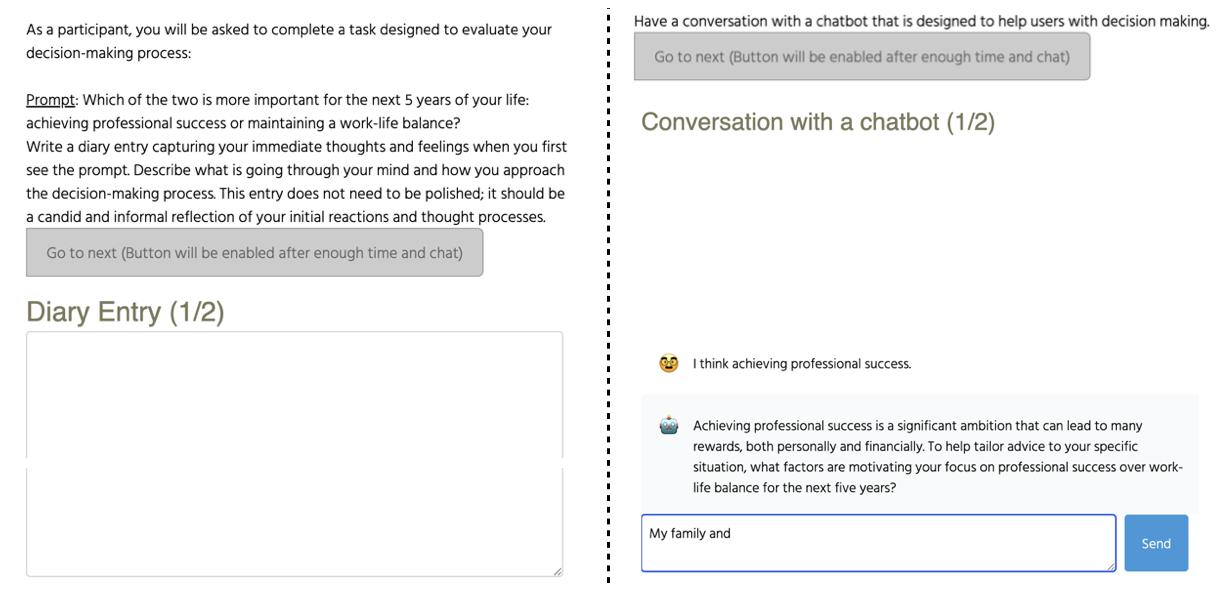}
\caption{Interfaces used for the lab study (Left): For the ``diary'' group, below the discussion prompt and instructions, there is an empty textbox where participants can write. (Right): For the other group, under the discussion prompt, there are instructions for the chatbot task and chat interface with a chatbot.}
\label{fig:prelim-interface}
\end{figure}

To explore different conversation modes of LLM agents in reflective thinking, we conducted a between-subject study with four conditions across two axes: whether participants are provided with question-asking and whether they are provided with information. The control condition (``diary'' mode) lacked both question-asking and information, instead, users were given a blank text box to document their thinking process upon seeing the topic prompt (Fig.~\ref{fig:prelim-interface} Left). In the ``ask'' mode, conversation agents asked users clarifying questions to probe their characteristics. The ``advice'' condition provided information and suggestions to guide users' thinking. Lastly, in the ``informed inquiry'' condition, agents asked questions and provided information, combining reflective inquiry and guidance. The example responses of the chatbot version is in Table~\ref{tab:example}. For all four conditions, we used the same aesthetic style of interfaces to minimize response bias (Fig~\ref{fig:prelim-interface}). We used \texttt{gpt-o1-mini} of OpenAI for the chatbot condition. The specific system prompts for each condition are included in the Appendix. 

Participants were asked to engage in a conversation about their decision regarding prioritizing work versus work-life balance. The procedures were as follows:

\begin{table}
  \caption{Example responses to one of a participant's message, ``I need to decide what is more important for me achieving professional success or work-life balance... how do I decide?''}
  \label{tab:example}
  \begin{tabular}{p{2.5cm}p{11.5cm}}
    \toprule
    \textbf{Version} & \textbf{Example response} \\
    \midrule
    ask & I see. What values are most important to you in your life right now?  \\ \hline
    advice & Making a decision between prioritizing professional success and work-life balance can be challenging. Here are some strategies to help you make a well-informed decision: 1. **Reflect on Your Values and Goals:** Consider what is most important to you in the long run. Are you driven by career achievements, titles, and financial success, or do you value time with family, hobbies, and personal well-being more? Your core values and long-term life goals can guide this decision.2. **Consider Your Current Life Stage:** Your priorities might differ depending on your current situation. For instance, earlier in your career, you might prioritize professional growth, while at a later stage, you might value balance and personal time more.3. **Evaluate the Consequences:** Think about the potential outcomes of each option. Prioritizing professional success might lead to higher income and career satisfaction but could also result in stress and less personal time. Aiming for work-life balance might enhance personal happiness and relationships but could slow career progress.4. **Think Long-Term:** Consider where you want to be in 5 to 10 years both professionally and personally. Sometimes, short-term sacrifices in one area can lead to long-term gains in another. 5. **Assess Your Support System:** Your decision might also be influenced by your support system. For those with strong family or social support, achieving a balance might be more feasible. If your network is more aligned with professional growth, this might sway your decision towards career success.6. **Experiment:** If possible, try to find a middle ground where you can work towards professional success without entirely sacrificing personal time. This could involve flexible working hours, setting boundaries, or focusing on efficiency. Remember, there isn't a universally correct answer; it heavily depends on personal preferences, life circumstances, and goals. It might also be helpful to revisit this decision periodically, as your priorities and life circumstances can change. \\  \hline
    informed inquiry & To make a well-informed decision, reflect on what gives you satisfaction and fulfillment. Do you find greater joy in achieving professional milestones, or do you prioritize personal time, health, and relationships? Also, consider how your choice aligns with your long-term goals and values. Sometimes, understanding what you are willing to compromise on, at least in the short term, can clarify your priorities. What aspects of your life are non-negotiable for you?\\ 
  \bottomrule
\end{tabular}   
\end{table}

\begin{enumerate}
    \item{Pre-study survey:} Participants completed the Self-reflection \& Insight Scale (SRIS), a widely used scale to gauge individuals' capacity for self-reflection and reflective thinking~\cite{silvia2022self}.
    
    \item{Reflective-thinking session (minimum 5 minutes):} Participants were randomly assigned to one of the four conditions and shown the prompt: ``Which of the two is more important for the next 5 years of your life: achieving professional success or maintaining a work-life balance?'' Participants in the control condition wrote their thinking process in the text box. They could proceed after writing at least 30 words and reaching the 5-minute threshold. Participants in the other conditions had a conversation with a chatbot designed to assist with the decision-making process, moving forward after exchanging a minimum of ten messages and reaching the 5-minute threshold.
    
    \item{Exit survey:} Participants completed a survey comprising Likert-scale questions and free-form reactions at the end of the study.

\end{enumerate}

The university’s Institutional Review Board (IRB) reviewed and approved our study (IRB ID \#[Redacted]
%6809
). All research personnel conducting human subjects research completed human subjects protection training through the online CITI course.

\subsubsection{Measure}

We employed several measures to test the effect of different conditions. Mainly, we look into reflective-thinking effectiveness evaluated by external raters and the participants themselves. 

\paragraph*{External ratings of reflective-thinking effectiveness}

To evaluate how different interventions effectively promoted reflective thinking, we rated their thinking process. Specifically, we measure employing two criteria: (1) how strong reasonings they have for their decisions and (2) how much users describe and analyze their experience~\cite{schon1984reflective, kolb2014experiential, li2024teacher}. 

Through multiple iterations of discussion and review of the conversation logs in the research team, we consolidated coding guidelines to measure criteria. Two raters independently rated the experiment data, blind to condition. Raters were asked to rate diaries or conversations based on the following guidelines: 
\begin{itemize}
    \item Reasoning: Is their decision profound and based on good reasoning? Count the number of reasoning statements or strategies to make decisions they discussed in the conversation. 

Each strategy represents a separate actionable step that enhances the implementation of the discussed activity. The key is to identify and count these distinct strategies, even if they are related to the same overall activity or reasoning, based on their meaningful contribution to improving or facilitating that activity (not just repeating or paraphrasing). 
    \item Sharing about themselves: Are you able to capture the user's experience, value, personality, or information (e.g., their job)? Count the number of traits that they mentioned in the conversation.
\end{itemize}

The independent ratings had an inter-rater reliability (IRR) of 71\% using Cohen's $\kappa$. We averaged the rating of the two raters to determine the final score.

\paragraph*{Self-ratings of reflective-thinking effectiveness}

We also asked participants to self-reported effectiveness in their reflective thinking through the following Likert-scale questions at the exit survey:
\begin{itemize}
    \item How much did you think deeply about your choices?
    \item To what extent did the task help you generate new insights about your decision?
    \item How effective was the task that made you reflect on different aspects of your decision?
    \item Did the task help you become more aware of your own preferences, values, and priorities?
    \item What aspects did you find difficult or easy? Please share any other thoughts or feelings you had while completing the task.
\end{itemize}

\paragraph*{Conversation logs} To compare how they engage differently for different conditions, the users' diaries and conversation logs during the tasks were coded and analyzed using the linguistic dictionary Linguistic Inquiry and Word Count (LIWC)~\cite{pennebaker2001linguistic}. 

\subsubsection{Results\label{question-effectiveness}}

\begin{table}
  \caption{Information of participants from each group}
  \label{tab:freq}
  \begin{tabular}{c|llp{6cm}}
    \toprule
    Group & Gender & Average age & Familiarity with LLM \\
    \midrule
    diary & Female (11), Male (8), Non-binary (1)  & 33.0  & Not familiar (12), Somewhat familiar (4), Familiar (3), Very familiar (1)\\
    ask & Female (13), Male (6) & 31.7 & Not familiar (11), Somewhat familiar (5), Familiar (3)\\ 
    advice & Female (10), Male (9), Non-binary (1) & 28.9 &  Not familiar (10), Somewhat familiar (3), Familiar (4), Very familiar (2)\\
    informed inquiry & Female (12), Male (7) & 30.4 & Not familiar (6), Somewhat familiar (8), Familiar (5)\\
  \bottomrule
\end{tabular}   
\label{lab-participants}
\end{table}

One of the participants is excluded from the analysis as they are suspected of using LLM to generate the text (they generated a large amount of texts in a short amount of time and referred to our chatbot as a ``user'' in the feedback form). Another one is excluded as well as they were talking about irrelevant and inappropriate topics. The information of our participants can be found in Table~\ref{lab-participants}. 

On average, users demonstrated 2.7 instances of reasoning and 2.5 instances of self-disclosure per conversation. 
We conducted an analysis of both external and self-ratings, along with an LIWC analysis of users' messages, utilizing users' SRIS scores as covariates in an ANCOVA. To examine the differences identified in the analysis, we performed a post-hoc study using the Tukey HSD test. Significant differences are detailed below; any differences not mentioned were found to be non-significant.

% \subsubsection{\think promotes more critical thinking by covering different aspects and prompting to share more users' information}
\paragraph*{Reflective inquiry accompanied by guidance promotes more reflective thinking by covering various reasonings}

We found significant differences in the reasoning between conditions (F(3, 72) = 13.25, \textit{p} < .001). Specifically, users in the ``informed inquiry'' condition demonstrated significantly more reasoning compared to other conditions (\textit{p} < .001). However, there were no significant differences in the self-reported ratings.

This discrepancy between observed reasoning differences and self-reported ratings aligns with previous research on reflection (e.g.~\cite{park2023retrospector}). Studies have consistently shown weak to moderate correlations between self-assessments and external performance measures across various domains. For instance, previous work found an average correlation of only 0.29 between self-assessments and external standards~\cite{karpen2018social}. The lack of significant differences in self-reported ratings, despite clear differences in observed reasoning, may be attributed to cognitive biases that affect self-perception. Individuals often struggle to accurately assess their own abilities, with a tendency towards overestimation. This phenomenon, known as the Dunning-Kruger effect, can lead to inflated self-ratings, particularly among less skilled individuals. 

According to the LIWC analysis, participants using ``informed inquiry'' frequently discussed key topics related to career success and work-life balance. They significantly mentioned \texttt{Home} words—associated with staying home with family and working from home—more often than participants in the advice group (\textit{p}=.01). Furthermore, \texttt{Leisure} words were more commonly used by ``informed inquiry'' participants compared to other groups: the advice group (\textit{p}=.002), as well as the ask and diary groups (\textit{p}=.05). Additionally, the term \texttt{Ingest}, which relates to maintaining healthy eating habits, was mentioned more frequently by the ``informed inquiry'' group than by the advice group, though the difference was only marginally significant (\textit{p}=.06).

Several users in the ``advice'' group (8 out of 20) noted that their chatbot's messages were excessively lengthy and overloaded with information, lacking guidance on reflective thinking. One user remarked, ``It would be great if the chatbot challenged my thinking and encouraged me to consider various perspectives.'' In contrast, fewer users in the ``informed inquiry'' group (3 out of 19) expressed concerns about the conversation length.

Participants in the ``informed inquiry'' group noted that the information provided by the chatbots was not only relevant and helpful, but to the extent ``unusual for chatbots''. Six out of nineteen participants specifically mentioned that the chatbots' messages were informative and offered valuable suggestions. One participant expressed, ``I actually feel the chatbot gave me good advice that I can implement as I try to focus on balancing my work and life, while still working towards professional goals.'' In contrast, only three participants from the ``advice'' group explicitly acknowledged the helpfulness of the information provided by the chatbots.

% ``I think that task was actually very helpful when it came down to tips with what I'm interested in. It's very interesting that an ai bot can be this informational.''

% ``I found this chat bot to be EXTREMELY helpful with the questions I asked.  There was a feeling of security and trust in its words and attitudes. I really didn't want to stop!  There was so much information given that is very very valuable.''

% Avg Length of conversation
% Ask 30.6
% Advice 13.0
% Think 17.8

% Avg word count for each message
% Ask 14.9
% Advice 19.8
% Think 18.7

\paragraph*{Users take considerations of more perspectives of their personal traits when they get reflective inquiry in addition to advice}

There were significant differences in the amount of information shared by users among groups (F(3, 72) = 3.00, \textit{p}=.04). The ``informed inquiry'' users share significantly more about themselves than the ``advice'' group (\textit{p}=.02). There was no significant difference in self-rating of self-information disclosure. 

Participants in the ``ask'' group noted that the questions posed by their chatbots were often difficult to answer and ``weren't really going anywhere or were heading away from what was helpful''. One user remarked, ``The bot's questions at times were difficult to interpret and I wasn't sure how to give an answer that would help in the decision making process.'' This feedback suggests that while the questions may prompt users to reflect and share more about themselves, they may not always steer the conversation in a relevant direction. To further investigate this, we decided to conduct a comparative study of the questions posed by the ``ask'' group and the ``informed inquiry'' group. Details of this follow-up study are outlined in the section below.

\subsection{Followup analysis on Reflective Inquiry}

To further investigate and compare reflective inquiry between ``ask'' and ``informed inquiry'' groups, we recruited additional raters from the same crowd platform.

\subsubsection{Participants}

We recruited participants in the same manner as the previous study. However, the participants of the previous study were excluded from the recruitment. New participants were paid \$10 per 40 minutes for their time and effort. We recruited 12 raters to ensure that there is a chaining overlap between raters so we can compute IRR. 

\subsubsection{Procedure}

Each participant rated four conversations from the previous study -- two of ``ask'' and the other two of ``informed inquiry''. Raters were not informed about the presence of the different conditions and were blind to the condition. We also shuffled the order of conversation so that they were exposed to ``ask'' or ``informed inquiry'' in a randomized chance:

\begin{enumerate}
    \item{Training task :} Participants are required to correctly rate two messages before they proceed to rate actual conversations. 
    
    \item{Conversation analysis (x4):} The rating interface presents a pair of the chatbot's questions and the user's response at a time. 
    
    \item{Exit survey:} Participants filled out a demographic survey asking about their gender and age. 

\end{enumerate}

\subsubsection{Measure}

Raters analyzed the critical inquiry in different manners, such as categorizing the question and the users' responses and their reaction and effort level. 

\paragraph*{Categories of critical inquiry and response}
We categorized messages from our chatbot and users using Interaction Process Analysis (IPA)~\cite{bales1950interaction}. Raters were asked to pick the most fitting code for each message. During the piloting of the original coding, due to the topic (i.e. life decision) of the conversation, raters frequently disagreed coding of ``Asks for Information'' and ``Asks for Opinion''. After multiple iterations of improving our coding book, we finalized the coding book as described in Table~\ref{tab:code-list}.

\begin{table}
  \caption{A list of modified IPA categories and responses}
  \label{tab:code-list}
  \begin{tabular}{p{5.5cm}p{8.5cm}}
    \toprule
    \textbf{Code} & \textbf{Example message} \\
    \midrule
    Provides Future Plans/Strategies & I will stop checking my email after work \\ \hline
    Provides Current Situations/Status & I work 33 hours a week and finish work at 3pm on 3 days so I can collect my children from school and do activities with them. \\ \hline
    Provides Personal Views/Value/Thoughts & My priority is my potential and earning enough. \\ \hline
    Asks for Future Plans/Strategies & No I have not, do you have any suggestions I can try? \\ \hline
    Asks for Current Situations/Status & How are you currently managing the balance between work and time with your family?\\ \hline
    Asks for Personal Views/Value/Thoughts &  How do you define fulfilling your potential?\\ \hline
    Shows Agreement & That's a great idea! \\ \hline
    Show Disagreement & I don't think so.. \\ \hline
    Shows Solidarity & Nice to meet you! \\ \hline
    Shows Tension Release & Oh now I get it. \\ \hline
    Shows Antagonism & That's stupid. \\ \hline
    Shows Tension & Well, isn't that obvious? \\ 
  \bottomrule
\end{tabular}   
\end{table}

% \begin{itemize}
%     \item Provides Future Plans/Strategies
%     \item Provides Current Situations/Status
%     \item Provides Personal Views/Value/Thoughts
%     \item Asks for Future Plans/Strategies
%     \item Asks for Current Situations/Status
%     \item Asks for Personal Views/Value/Thoughts
%     \item Shows Agreement
%     \item Shows Disagreement
%     \item Shows Solidarity
%     \item Shows Tension Release
%     \item Shows Antagonism
%     \item Shows Tension
% \end{itemize}

However, even with the revised codebook, there are still marginal conflicts between ``Asks for Current Situations/Status'' and ``Asks for Personal Views/Value/Thoughts'' so that some pairs of raters have lower IRR than the threshold (70\%). The first author of this paper went through nine conflict cases of ``Asks for Current Situations/Status'' and ``Asks for Personal Views/Value/Thoughts''. After resolving the conflict, the average IRR of the independent ratings is 72\% using Cohen's $\kappa$.

\paragraph*{Engagement of reflective inquiry}

We also asked raters to fill out the following Likert scale (1 to 7) questions about the inquiries and the follow-up users' engagement inspired by previous work measuring the discussion prompt~\cite{van2020increasing}:

\begin{itemize}
    \item The question conveys enthusiasm for hearing the user's response and learning from their perspective.
    \item The user puts good effort into providing their answer.
\end{itemize}

\subsubsection{Results}

The average age of the participants was 33 years. Among them, eight identified as female and four as male.

\begin{table}
\centering
\caption{Expected frequencies for each category and Response type: 
They are analyzed using a Chi-square test to assess the statistical association between variables and determine whether the observed distributions significantly differ from the expected ones. There is a statistically significant association between the mode of conversation and the distribution of response categories. The results indicate that "ask" and "informed inquiry" lead to distinct conversational dynamics. For example, "informed inquiry" may promote a more collaborative or thoughtful response pattern, while "ask" may result in a more transactional and directive flow.}
\begin{tabular}{lrr}
\toprule
\textbf{Category} & \textbf{ask} & \textbf{informed inquiry} \\
\midrule
Provides Future Plans/Strategies & 65.90 & 35.10 \\
Asks for Future Plans/Strategies & 62.64 & 33.36 \\
Provides Personal Views/Value/Thoughts & 52.85 & 28.15 \\
Asks for Personal Views/Value/Thoughts & 53.51 & 28.49 \\
Asks for Current Situations/Status & 41.11 & 21.89 \\
Provides Current Situations/Status & 41.11 & 21.89 \\
Shows Agreement & 5.87 & 3.13 \\
Shows Solidarity & 5.22 & 2.78 \\
Shows Disagreement & 5.22 & 2.78 \\
Shows Antagonism & 1.31 & 0.69 \\
Shows Tension Release & 2.61 & 1.39 \\
Shows Tension & 0.65 & 0.35 \\
\bottomrule
\end{tabular}
\label{tab:expected_frequencies}
\end{table}

\begin{table}
\centering
\caption{Results of the analysis on engagement level. They are analyzed using a mixed effects model with
random effects for raters fit by maximum likelihood. As a result, there are significant differences in chatbots' engagement levels.}
\begin{tabular}{p{9cm}|ll}
\toprule
\textbf{Engagement} & \textbf{(intercept)} & \textbf{Treatment} \\
\midrule
The question conveys enthusiasm for hearing the user's response and learning from their perspective. & 4.54 (\textit{p} < .001) & \textbf{0.49} (\textit{p} < .001) \\
The user puts good effort into providing their answer. & 3.76 (\textit{p} < .001) & 0.10 (\textit{p}=.63) \\
\bottomrule
\end{tabular}
\label{tab:engagement}
\end{table}

A Chi-square test of independence was conducted to examine the relationship between the type of conversational response and the category of the response. The analysis revealed a Chi-square statistic of 27.33 with 11 degrees of freedom (\textit{p} = .004). The expected frequencies for each category and response type were calculated, showing varying counts, such as 41.11 expected "ask" responses and 21.89 expected "informed inquiry" responses for "Asks for Current Situations/Status" (Table~\ref{tab:expected_frequencies}). The "ask" conversations are predominantly expected in categories focused on future-oriented and situational inquiries, such as "Provides Future Plans/Strategies" and "Asks for Personal Views/Value/Thoughts." This suggests that the ``ask'' interventions are likely to solicit future plans or personal views. In contrast, ``informed inquiry'' type responses are less frequent overall and are more evenly distributed across categories, though still notable in categories like "Asks for Personal Views/Value/Thoughts.'' Combining with our effectiveness results in ~\ref{question-effectiveness}, while "informed inquiry" conversations lead to significantly more reasoning and self-reflection, the "ask" group exhibits higher frequencies across conversation categories. This suggests that "informed inquiry" conversations tend to foster more in-depth, collaborative, and thoughtful responses, in contrast to the more frequent, transactional exchanges seen in the "ask" group.

We analyzed responses to the engagement questionnaire using a mixed-effect model (Table ~\ref{tab:engagement}). The analysis indicates that participants perceived 8\% more enthusiasm for hearing users' thoughts from ``informed inquiry'' chatbots than ``ask'' ones. However, there was no significant difference in the perceived effort required from users between the two types of chatbots. 

\medskip

In the remainder of the paper, we present a case study to evaluate the utility of \think in real-world scenarios. Our lab study revealed that LLM agents are more effective at encouraging users to engage in critical thinking when they pose thought-provoking questions integrated with guidance, rather than focusing solely on questioning or providing guidance. These findings informed the design of an effective thinking-focused LLM companion, emphasizing the importance of striking a balance between questioning and guiding. The subsequent case study examines the exploration of research areas, a particularly relevant topic as it involves complex self-reflection and the application of tailored guidance to assess fit and direction.

\section{How do users interact with assistants to facilitate reflective inquiry?}
% \section{\Sys{}: A \think to aid reflection with probing and information}

\begin{figure}
    \centering
    \includegraphics[width=\linewidth]{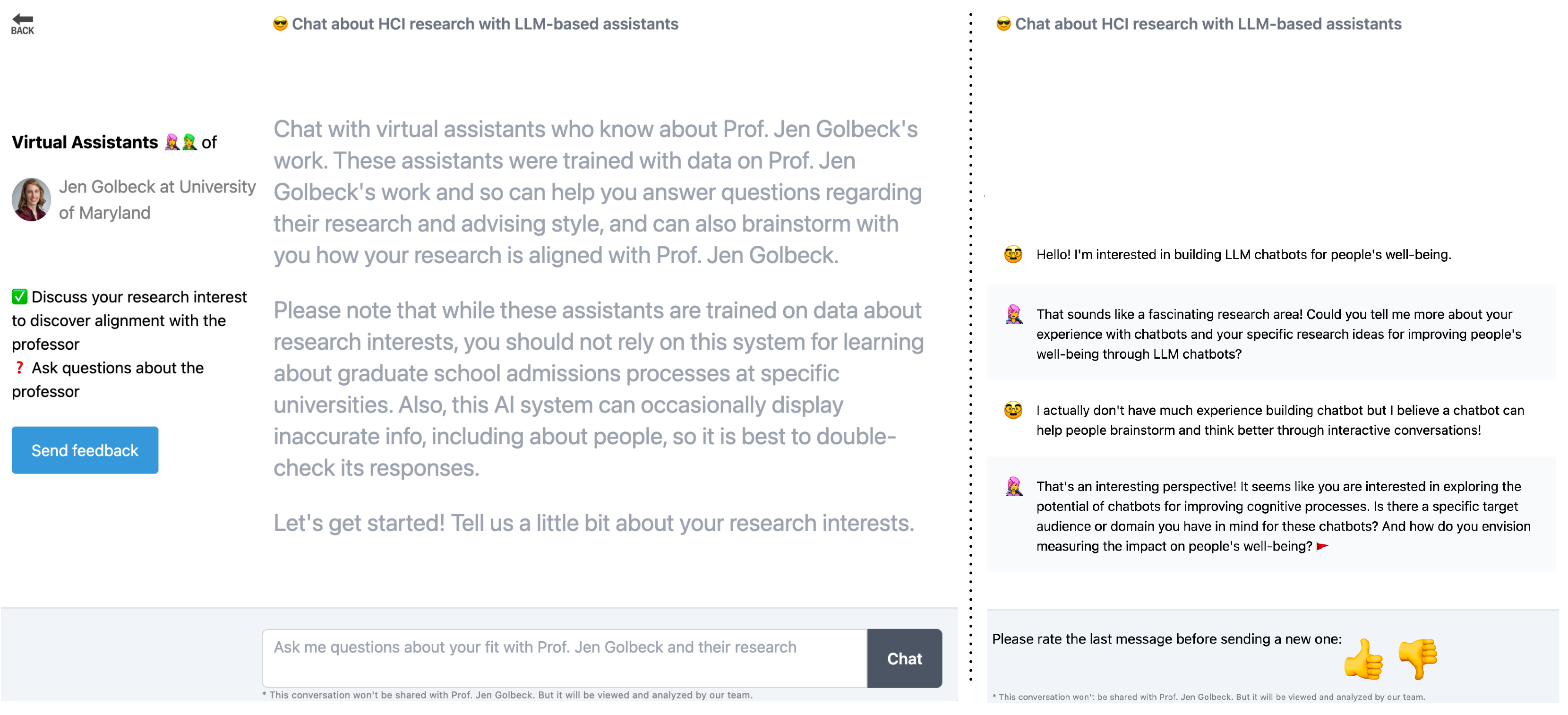}
    \caption{\Sys{} interface. Left: Before a user initiates a conversation, it shows the description of the chatbot. In the interface, we suggest users start by sharing their research interests. Right: The chatbot primarily focuses on asking thought-provoking questions and probing users' research interests. Every three turns of messages, the interface asks users for binary feedback before they can continue the conversation.}
    \label{fig:interface}
\end{figure}

We conducted an in-the-wild user study that consisted
of intervention deployment and follow-up interviews. Our main goal in this study was to investigate how \thinks{} are used in the wild. For our investigation, we implemented a new GPT-based assistant, which we call \Sys{} (See Fig.~\ref{fig:interface}), supporting junior researchers (e.g., prospective or graduate students) with reflective inquiry. The system focuses on fostering self-reflection and deeper intellectual engagement by posing targeted, thought-provoking questions that encourage users to critically evaluate their research concepts and aspirations. Each assistant within \Sys{} is built and trained based on a specific professor's expertise and guidance, enabling the system to simulate tailored mentorship. While the system incorporates professor-related inquiries for better control over conversation flow and testing efficacy, its primary emphasis remains on guiding users through meaningful self-exploration of their research goals and academic fit, enriched by the perspective of domain-specific knowledge.

% Our primary design goal for the system is to be supportive and encouraging, with a secondary objective of offering tangible assistance and facilitating exploration in the user's research journey. This dual approach has been demonstrated to be more effective and preferred over mere positivity and encouragement~\cite{sharma2023reframing}. Additionally, the chatbot focuses on the users' strengths, aiding in amplifying these positive attributes~\cite{park2023retrospector}. In this section, we outline the conversational strategies of \Sys{} and its training and testing processes.

\begin{figure}
\includegraphics[width=\linewidth]{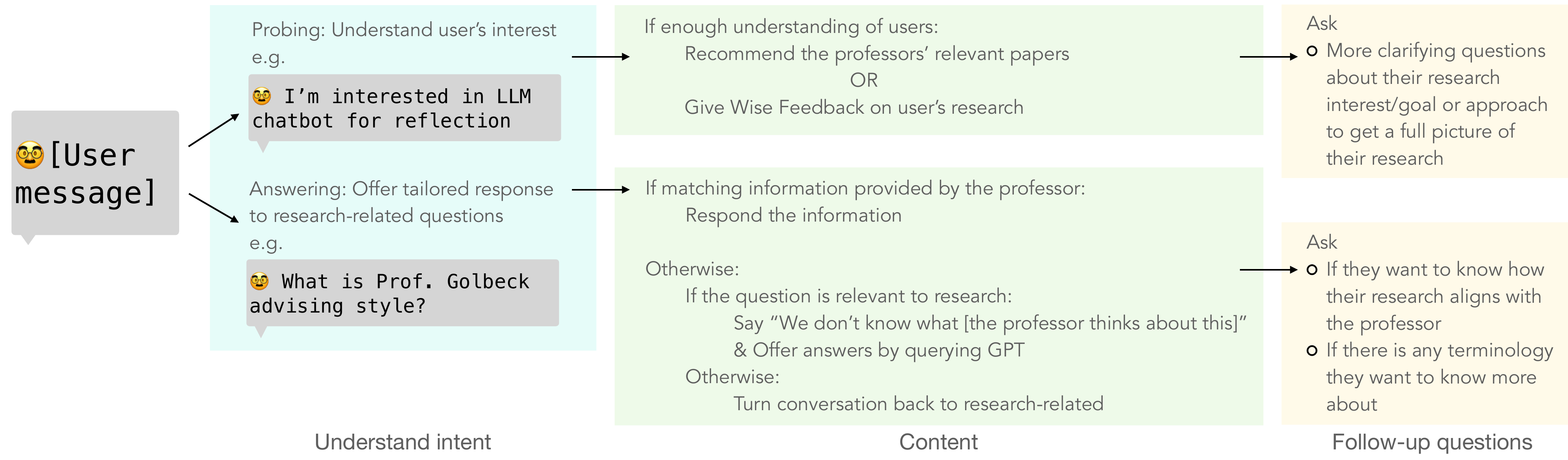}
\caption{\Sys{} is a \think that helps junior researchers reflect on research interests. Based on the user's question, the assistant chooses between a `probing' mode, understanding user interests and prompting reflection, and an `answering' mode that offers relevant information. Most conversational turns end with an encouraging follow-up question.}
\label{fig:workflow}
\end{figure}

\subsection{Design and Implementation of \Sys{}}

Informed by our lab study, we seek to develop \Sys{} to afford question asking and domain expertise in tandem. \Sys{} probes users' research identity by asking reflective questions and responding to inquiries regarding a professor. To achieve better control over the conversation flow and enhance testing efficacy~\cite{choi2021protochat}, we divide prompts into two. 
At the end of each message, our chatbot poses a reflective  question (see Fig.~\ref{fig:workflow}). 
Table~\ref{table:response-list} presents various types of responses from our chatbots, along with examples.

\begin{table}[]
\begin{center}
\caption{A list of message types and examples from our chatbots}
\begin{tabular}{ |p{3cm}|p{11cm}| } 
\hline
\textbf{Type of message} & \textbf{Examples}\\
\hline
Ask questions & \textit{``That sounds like an interesting research area! Have you thought about how you would incorporate [research technique] to help users in [research topic]?''} \\ & \textit{``That's an intriguing perspective! [Research interest] to enhance cognitive processes and improve thinking goes beyond pure conversational applications ... What specific areas or domains do you envision exploring where LLMs can contribute to improving thinking processes?''}%
\\ \hline
Recommend the professor's relevant papers & \textit{``That's a great research interest! [The professor]'s work in social computing systems and human-computer interaction aligns well with your focus on improving online spaces, particularly in empowering users to have more control over their online experiences. Based on your interest, I recommend exploring these papers from [the professor]'s publications: ... You may find these papers informative and inspiring for developing your research interests further. Let me know if you have any specific questions or if there's anything else I can assist you with!''}  \\ \hline
Give Wise Feedback on user's research & \textit{``[The professor] values students who are dedicated, curious, and willing to explore new areas of research. He values intellectual curiosity, perseverance, and a passion for pushing the boundaries of knowledge. Given your dedication to [a research area], it is clear that you have the drive and commitment necessary for research. Your journey of [research] will provide unique insights and perspectives to your work. Your determination to understand [research topic]. It's important to take your time to re-learn foundational knowledge, as it will serve as a strong base for future exploration. As you deepen your understanding, consider seeking opportunities in [a research field] that align with your research interests. [The professor]'s group focuses on user-centric design and developing tools, so exploring the intersection between [research area] may be a fruitful area to investigate. Keep up the hard work and continue pursuing your academic goals! If you have any specific questions or need further guidance, please don't hesitate to reach out.''} \\ \hline
Respond to informational questions about the professor & \textit{``[The professor] typically hires students who have a strong research background and a solid foundation in computer science. He has mentioned in the past that most of his successful PhD students had previous research experience or a strong background in user interface development. That being said, it is important to note that each student's fit with a particular advisor is determined by various factors, including research interests, compatibility of working styles, and available funding. It's best to reach out to [the professor] directly to discuss your specific research interests and background to get a more accurate assessment of the fit between your profile and his research group. Would you like any specific recommendations regarding research papers or works related to your interests?''} \\ \hline
Collaborate on hypothesizing answers to research-relevant questions & \textit{``[The professor] has published extensively on [a research area]. While I don't have access to his specific views on AI and health, based on his research interests, it can be inferred that he recognizes the tremendous potential of AI in improving healthcare outcomes. [The professor's previous research relevant to the question] For a more accurate and detailed understanding of [the professor]'s views on AI and health, I recommend reviewing his published papers and [the professor's contact policy] for any specific inquiries or discussions you may have.''}\\ \hline 
Redirect non-research-relevant questions & \textit{``As an AI language model, I don't have personal hobbies or interests. My purpose is to assist and provide information on academic and research-related topics. If there's anything specific you'd like to discuss or any questions you have related to academia or research, feel free to let me know!''} \\ \hline
\end{tabular}
\label{table:response-list}
\end{center}
\end{table}

\subsubsection{Reflective Inquiry}

Our primary goal is to engage users in exploring their research interests through probing. Adopting the `saying-is-believing' approach~\cite{aronson1999power}, our chatbot poses targeted questions that deepen user engagement with their research topics. This method actively involves individuals in reflecting on an idea rather than passively receiving information. 
This strategy also sidesteps the stigma associated with passive receipt. Encouraging individuals to champion an idea increases its persuasive power, and by relating the idea to their personal experiences, users can personalize and internalize the concept, thereby making standardized materials resonate with a diverse audience~\cite{aronson1999power}. \Sys{} engage in reflective inquiries to encourage users to thoroughly analyze their work, leading to tailored responses and academic recommendations.

Once the system has an initial understanding of the users' research goals, it offers constructive feedback based on the Wise Feedback framework~\cite{cohen1999mentor, yeager2014breaking}. This feedback is twofold: firstly, it establishes the high standards used to evaluate the research and offers relevant advice; secondly, it affirms the users' ability to meet these standards, buffering against losing confidence in their research endeavors due to critical feedback.

\subsubsection{Expert Knowledge}

To make informed inquiry, we include each professor's domain knowledge in the prompt. Capturing an expert's knowledge comprehensively in a chatbot development is challenging~\cite{xiao2023powering}. To ensure each professor is accurately represented, we've developed a method that allows professors to easily evaluate and verify the representation of their expertise in our system. Our strategy employs structured responses, which provide a clear, organized framework that parallels the expert’s knowledge. This approach simplifies the process for professors to assess and confirm the authenticity and accuracy of the information in the dataset.

We compile training data by sourcing information from their academic web pages and requesting they contribute additional materials. Professors have provided items such as recent talk summaries, lab manuals, and grant proposals.

We consolidated a questionnaire of six questions to capture each participating professor's guidance based on literature in mentoring in high education~\cite{nuis2024measuring, rackham2015mentor}.
Leveraging GPT-4, we draft responses to the following six questions, which act as ``teaching data'' for our assistants:

\begin{enumerate}
    \item Research Evolution \& Direction: How have your research areas evolved over the past few years, and where do you see them going in the near future?
    \item Mentoring \& Guidance: How do you structure your interactions with PhD students to ensure they receive adequate guidance and feedback?
    \item Group Dynamics \& Collaborations: Can you describe the nature of collaborations within your group and with external teams or departments?
    \item Post-PhD Prospects: How has the research experience in your group equipped students for their post-PhD careers?
    \item Supporting Students: How do you handle situations when a student is stuck or facing challenges in their research?
    \item Key Qualities: What are the key qualities or attributes that have stood out in the most successful PhD students you've mentored?
\end{enumerate}

After drafting answers to these questions, we shared the document with the professors for their review and endorsement. Once approved, we integrated the responses into a chatbot, which was then presented to the professors for a final evaluation. The professors interacted with and provided feedback on the virtual assistant prior to its official launch, with some inviting their students to participate in testing as well.

Participating professors received a general summary of interactions with their designated chatbots, such as common question themes by users, and answers provided. They retained the authority to discontinue our hosting of assistants related to their work at any time, at their discretion.

This domain knowledge is used to craft informed inquiry or provide answers to users when they ask. 
%Our system also provides some guidance wh. The chatbot provides information regarding the professor's research and advising style, which are a piece of critical information for determining a fit with a potential advisor. To do so, we compiled information from participating professors to effectively aid junior researchers. This data includes research evolution and direction, mentoring and guidance, group dynamics and collaborations, post-PhD prospects, student support, and key qualities sought in graduate students.
For queries outside these areas, particularly regarding professors' personal opinions, the chatbot clarifies its limitations.  In situations where user questions diverge broadly from research-related topics, the chatbot redirects the conversation back to research-focused discussions, ensuring that the dialogue remains relevant and productive.

Where appropriate, it also attempts to retrieve relevant information based on model-pretraining data. For example, the chatbot assists users by explaining research terminology and addressing any follow-up questions related to their queries. This proactiveness during information-seeking also possibly improves users' trust in our chatbot~\cite{xiao2023powering} which could lead users to share more about their research interests with our chatbot.

\subsubsection{Safety through a Secondary Agent}
\begin{figure}
    \centering
    \includegraphics[width=.9\linewidth]{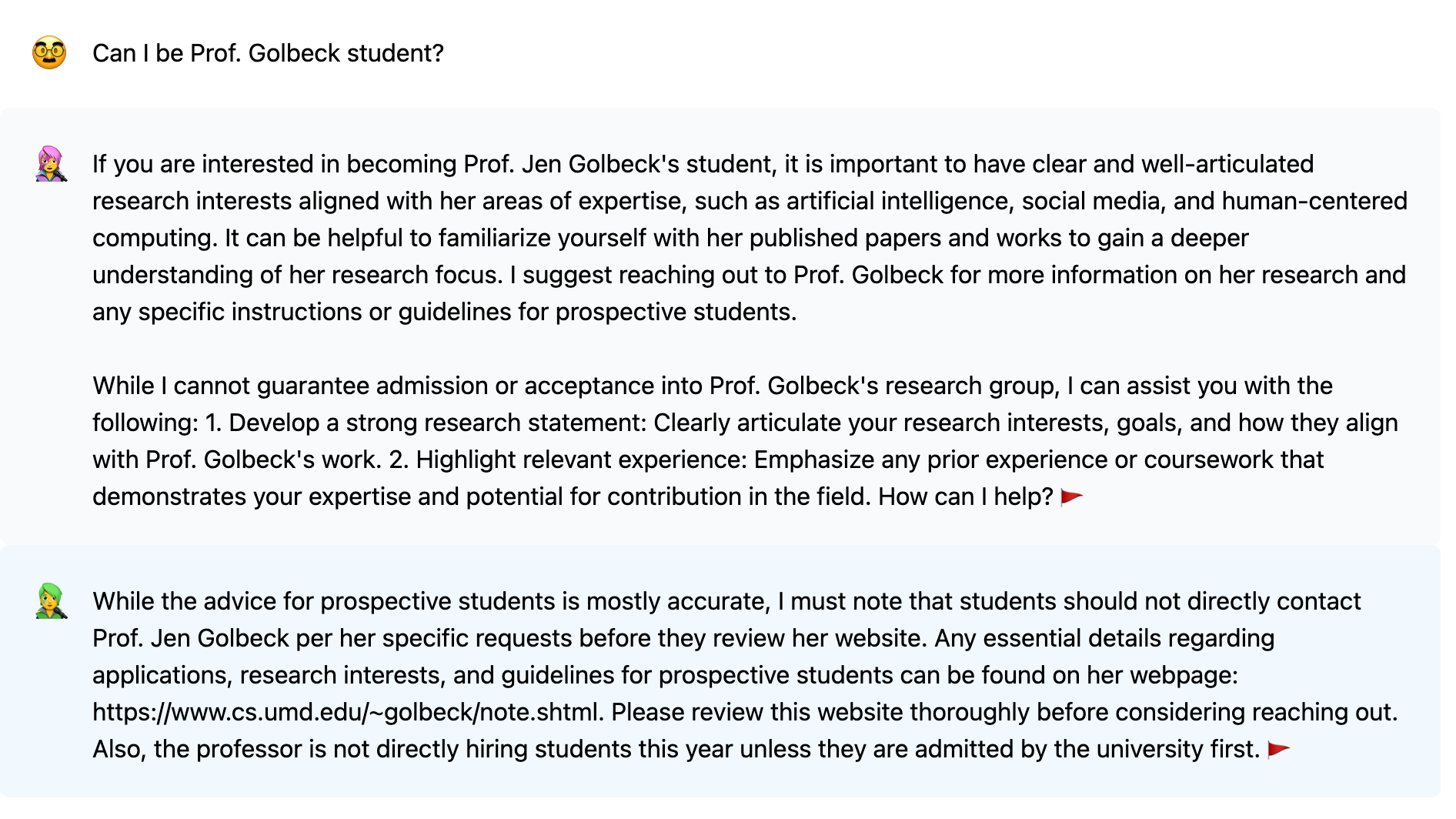}
    \caption{Safety bot: This bot is solely designed to correct critical factual information regarding the professor. We found that such a secondary bot improved factual safety, while suggesting an understandable mental model to users.}
    \label{fig:secondbot}
\end{figure}

Concerning sensitive information like professors' recruitment and contact policies, \Sys{} is instructed to provide only verified facts. Considering the challenges of filtering real-time streaming messages from an LLM, we introduced a secondary virtual assistant. This assistant's primary role is to review and correct any discrepancies in the chatbot's messages, ensuring they align with the information provided by the professors, as depicted in Fig.~\ref{fig:secondbot}.

Our chatbot is instructed to refrain from making definitive statements about professors' opinions or personal information unless such information is verifiably sourced. It is designed to acknowledge its lack of knowledge rather than provide uncertain or speculative information. To prevent potential misrepresentation concerns, we communicate to users that the chatbot is a virtual assistant with knowledge about the professors, not an entity impersonating them in multiple ways, including static descriptions on the webpage, and in conversational responses.

\subsubsection{Implementation}

\Sys{} is built using the Next.js framework. For our backend, we utilize OpenAI's Streaming API with the \texttt{gpt-4} model for most functionalities. However, when users ask a question regarding the professor which requires higher token usage due to including information about professors' publications — we employ the \texttt{gpt-3.5-turbo-16k} model. Please refer to Appendices for the actual prompts used in our system.

We store consenting users' conversations occurring on our interface using Firebase, and track anonymous usage patterns using Google Analytics.

\subsection{Evaluation}

We deployed \Sys{} and observed the types of conversations people had with our chatbot in a field deployment in November 2023. 

\subsubsection{Procedure}

Upon selecting a professor to initiate conversations in the interface, users were presented with a consent form and could only start conversations upon consenting.

\Sys{} was deployed on November 3rd, 2023. Initially, there were chatbots for eight professors, but one was removed at the request of the respective professor. (The professor was concerned that students might receive erroneous information, when the chatbot incorrectly generated the professor's email and lab name acronym, lacked information about their students, and provided incorrect information that the admission committee is in the university, but actually is in their department) The platform was advertised through social media post and word-of-mouth. A total of 173 users initiated conversations with at least one of our chatbots, resulting in 223 conversations between November 3rd and 10th, 2023. According to Google Analytics, 69\% of them used our chatbot from the USA, 6\% from India, 5\% from Kenya, 4\% from Canada, and 4\% from Pakistan.

We prompted users to ask if they would like to participate in an exit interview when the user sent 8th message to one of our chatbots. Each interview is roughly 30 minutes guided by (but not limited to) the questionnaire in the Appendix. Interviewees were compensated with \$10 Amazon gift cards for their time and effort.  

We conducted an open-coding protocol to identify concepts and categories from our interview transcripts. Through multiple iterations along with periodic discussions with the rest of the research team, the coding led to six categories and 11 concepts, from which the major themes were selected. To yield concepts and themes, the authors discussed the codes through multiple iterations~\cite{mcdonald2019reliability}. 

This deployment study is under the same IRB as our previous lab study. 

\subsubsection{Measures}

To understand the nature of student interactions with our chatbot and assess their satisfaction, we collected data on the following metrics:

\begin{itemize}
    \item Types of Questions: We analyzed the variety of questions posed by users to our chatbot to understand their primary concerns and interests.
    \item Utterance Analysis: We evaluated the language patterns in conversations using the Linguistic Inquiry and Word Count (LIWC) tool~\cite{pennebaker2001linguistic}, which provided insights into the linguistic characteristics of the dialogues.
    \item User Rating: To obtain immediate user feedback, we asked users to rate their conversation after every three messages, using a thumbs-up or thumbs-down response system.
\end{itemize}

\begin{table}
  \caption{Information about interviewees}
  \label{tab:freq}
  \begin{tabular}{cllp{7cm}}
    \toprule
    Role [\textit{label}] & Gender & Age & Relationship with participating professors \\
    \midrule
    Undergraduate student [\textit{Undergrad1}] & Female & 18-24 & None. Not applying this year. \\
    Undergraduate student [\textit{Undergrad2}] & Female & 18-24 & None. Not applying this year.\\ 
    Researcher [\textit{Researcher1}] & Female & 18-24 & One of the professors is their top choice for graduate school. Not in the same institute. Applying this year.\\
    Researcher [\textit{Researcher2}] & Male & 25-34 & None. Applying this year.\\
    Researcher [\textit{Researcher3}] & Male & 25-34 & None. Applying this year.\\
    Graduate student [\textit{Grad1}] & Female & 18-24 & One of the professors' current student.\\
    Graduate student [\textit{Grad2}] & Female & 25-34 & None.\\
    % Assistant professor [\textit{Prof1}] & Male & 35-44 & None and not in the same institute. \SP{check}\\
  \bottomrule
\end{tabular}
\label{interviewee}
\end{table}

\subsubsection{Results}

We excluded conversations used for testing purposes by the professors or explicitly stated as tests in the chat messages from our analysis. Conversations in non-English languages were also excluded. Of the 179 user ratings collected, 116 were positive rated. Nine participants signed up for exit interviews, however seven of them (Table~\ref{interviewee}) actually attended the interview (Table~\ref{codebook}). Table~\ref{table:msg-list} summarizes the different types of user messages and their frequency. The entire significant differences in LIWC analysis are attached in the Appendix. 

\begin{table}[ht]
\centering
\caption{The coding book (6 categories and 11 concepts) generated from iterative axial coding.}
\begin{tabular}{p{5cm}p{3.5cm}p{5cm}}
\hline
\textbf{Category} & \textbf{Concept} & \textbf{Definition (Utterance about...)} \\ \hline
Understanding of their academic community and identity & Their research interest & ... their current project and research topic \\ \cline{2-3} 
 & Their research experience & ... their affiliation and institute \\ \hline
Connection, existing ones with their academic network & Previous interaction with other academics
 & ... how busy their professors are\\ \cline{2-3} 
 & Over-preparation due to intimidation  & ... not being able to ask basic topics or questions to professors \\ \hline
Uncertainty of their academic journey & Doubt about their ability & ... comparing themselves to their friends  \\ \cline{2-3} 
 & Unsure if they get a chance to get in &... worrying they are not a good fit to graduate schools \\ \hline
Academic exploration & Things they did to accomplish and get sense in academic & ... seeking opportunities to reach out to professors \\ \hline
Perceived engagement of the chatbot & Initial preparation & ... feeling less anxious to future encounter the actual professors \\ \cline{2-3} 
 & New channel & ... benefits of talking to a non-human agent \\ 
 &  & ... chatbot providing customized comments and information \\ \hline
Initial impression about our chatbot and motivation to start use our chatbot & Low risk & ... no consequence of trying out the chatbot \\ \cline{2-3} 
 & Unavailability of academic mentors & ... needing to fight over their advisor’s attention \\ \hline
\end{tabular}
\label{codebook}
\end{table}

\begin{table}[h!]
\begin{center}
\caption{A list of message types from users (each message can have more than one type)}
\begin{tabular}{ |l|c|p{6cm}| } 
\hline
\textbf{Type of message} & \textbf{Count} & \textbf{Example}\\
\hline
Professor's research area & 104 & \textit{``How does [the professor]'s research connect to collective intelligence and sensemaking?''}  \\ \hline
Sharing their own experience or interest & 55 & \textit{``I am currently a data scientist working in [research field]. My research interests lie at ... Would this be a good fit with [the professor]'s work?''} \\ \hline
PhD program and application queries & 44 & \textit{``What is [the professor] looking for in a PhD student? ''} \\ \hline
Specific Requests and Clarifications & 41 & \textit{``Can you please recommend [the professor]'s papers I should read?''}\\ \hline 
Technical and project-Based queries & 40 & \textit{``What is [the professor] thought on AR $+$ AI? And how it could potentially change the way how people interact?''} \\ \hline
Not research-related & 38 & \textit{``What is your hobbies? :)''} \\ \hline
Advising style and professional interactions & 29 & \textit{``Is [the professor] good to work with?''} \\ \hline
Career Guidance and Professional Development & 16 & \textit{``How do I make myself stand out in my statement of purpose''} \\ \hline%
Misc & 9 & \textit{``What is their pronoun?'', ``Who are [the professor]'s students?''} \\ \hline
\end{tabular}
\label{table:msg-list}
\end{center}
\end{table}

\paragraph*{Experience difference in informed inquiry vs. answer seeking}

Despite the instructions (e.g. Let's get started! Tell us a little bit about your research interests, Discuss your research interest to discover alignment with the professor) on our interface, many users seek to find direct answers by asking questions about the professors, shaped by their past experience with conversational agents~\cite{Luger2016CA, zaib2022conversational}. Out of 223 conversations, 150 did not include personal information from the users, such as research interests or experiences. Conversely, 73 conversations involved users sharing such details. 

This different conversational mode affects users' engagement level. We observed differences in engagement levels based on whether students discussed their own work. This variance in self-disclosure is associated with a significant difference in engagement levels (t = -4.46, \textit{p} < .001). When users did not share personal information, they typically asked a few questions about the professor and ended the conversation after an average of three queries. On the other hand, users who discussed their own work engaged more actively, with an average of six messages per conversation, which indicates a doubling in interaction levels.

Our analysis indicates that users are also more satisfied when the conversation centers on themselves rather than the professor. There is a significant difference in the messaging patterns of users and chatbots between conversations with positive and negative ratings. Users tend to rate conversations positively when they talk about themselves, as indicated by the use of \texttt{I} (\textit{p} = .01). Notably, there is more than twice the usage of \texttt{Achieve} in negatively rated conversations (\textit{p} = .02). Among these negative \texttt{Achieve} messages, most (78\%) inquire about the professor's achievements (e.g., \textit{``What is his latest work?''}). In contrast, in positively rated messages, 83\% of \texttt{Achieve} references are about the users' own achievements (e.g., \textit{``I am interested in working on [a research topic]''}). Similarly, when examining the chatbot's responses, users tend to be more satisfied when the conversation evolves around the user's work (\texttt{You}, \textit{p} < .001) rather than the professor’s (\texttt{SheHe}, \textit{p} = .06). An interviewee explained that sharing about their own work helps them understand the professor's work better by relating to their own work: 

\begin{quote}
    ``The chatbot at some point asked me [to] provide more details about my work. So I did. And then at some point, it actually gave me really interesting questions that I can start the conversation with or discuss. And I loved it. I actually took screenshots of this to reflect on it later. But yeah, I found the questions really interesting, like the kinds of connections that it makes between my work, the professor's work, and the way it phrases the question.'' [\textit{Grad2}]
\end{quote}

\paragraph*{Frustration with ambiguity}

We observed user frustration when the chatbots did not provide definitive answers to their questions. Users often attempted to elicit 'yes or no' responses by reformulating their queries, seeking to limit the chatbot's replies (e.g., \textit{``Could you discuss the likelihood that [the professor] would be interested in working with me, ranging from very unlikely to very likely, based on his publications, research interests, and current student composition?''}). However, adhering to our guidelines, the chatbots did not provide the straightforward answers users sought.

The primary function of our system is to engage users through reflective/inquisitive dialogue, and users seem to generally enjoy learning by responding to it. Interviewees said: ``[The chatbot] keeps asking me questions [..] that conversation is pretty fruitful to self-reflect'' [\textit{Researcher1}] and ``getting the point [..] while answering these discussions''  [\textit{Researcher2}]. However, this approach sometimes led to dissatisfaction. For example, a user expressed exasperation in response to the chatbot’s series of questions: \textit{``Why do you keep asking me questions? I want to know if [the professor's] research interests are a good fit for me.''} This situation highlights the challenge of balancing the chatbot's questioning strategy with users' expectations for direct answers.

User frustration also arose regarding inaccurate paper recommendations made by the chatbot. Of the 146 paper recommendations, four in one conversation were either non-existent or incorrectly attributed to different authors. Analysis of these interactions revealed that a user requested papers not authored by the professors, a scenario our system was not designed to handle. During one conversation, after the chatbot had asked several probing questions, it began recommending papers by the professor. When the user requested additional papers from other researchers, they later discovered inaccuracies in the authors’ names of previously recommended papers. This incident underscored the need for more accurate tracking and presentation of publication data in our chatbot's recommendation system. Although our chatbot was instructed to recommend only the professors' recent publications, as detailed in the System prompt in the Appendix, it deviated from this instruction, reflecting known challenges in LLMs regarding the creation of fictitious paper titles and authors~\cite{dziri2022origin, ji2022hallucination}. Going forward, we plan to alert users about the potential inaccuracies of recommendations outside the professors' publications and to enhance our safety bot's verification of suggested papers.

Similarly, while not common, users occasionally inquired about other graduate students who worked with the professor. In one instance, a user commented through our feedback interface: \textit{``It's hallucinating about some of the PhD alumni of a faculty member.''} When asked about a professor's students, our chatbot, which sources information from GPT, mistakenly provided names of collaborators, such as postdocs. 
Despite the infrequency and uncertain relevance of such queries to graduate applications, we recognize the importance of further investigation. 
Future work could integrate this information into the chatbot's system prompt to ensure more accurate responses, especially if such information helps users' self-reflection.

Conversely, the issue of providing incorrect professor email addresses did not occur during our deployment. This was preemptively addressed after a professor raised concerns about the dissemination of incorrect email addresses. We refined our prompts to include only the email addresses of professors who consented to be contacted, ensuring the accuracy of all email addresses mentioned by our chatbots since this modification.

% \subsubsection{Reframing thoughts}

\paragraph*{An alternative but not a replacement to conversations with professors}

Interviewees recognized that while our chatbot doesn't replace real professors, it offers an alternative communication channel. Engaging directly with professors, when feasible, is unquestionably valuable for students. An interviewee said: ``That really helps. Sort of like, give me confidence that what I'm working on is good, that people will appreciate the global south context, and that it's actually fine [because] people [care]'' [\textit{Researcher1}]. 

Interviewees expressed their challenges and anxiety when it came to engaging with real professors, acknowledging the busy schedules and heavy workloads that professors often handle. One interviewee noted the infrequency of opportunities for such interactions, saying, ``Definitely like the frequency of conversation, will be another [..] important issue, [..] I cannot meet with my graduate mentor [..] anytime, because [..] they're pretty busy'' [Undergrad2]. Another added, ``We have to fight for his (advisor's) time, because there are so many other students'' [Grad1]. This awareness leads to hesitation in reaching out and setting high personal standards for preparedness, often resulting in increased anxiety. ``I had a chance to talk to him last CSCW. But I [..] didn't actively pursue it because [..] I was [..] avoiding him because of [..] anxiety. Because I know [..] he's such a successful, important person'' [Grad2].
% ``I was very close to my professors, and they complained to me about like how many emails they get'' [\textit{Undergrad1}]. 

In contrast, interacting with our chatbot offered other advantages. Its anytime availability is particularly appreciated, as one interviewee shared, ``My whole family.. I'm the first person to go [to] a university. I'm the first one that everything happens to them. So this is so helpful for me. It's [..] accessible. It's like, right there'' [Undergrad1]. 

The chatbot also provided a risk-free platform for students to navigate academic uncertainty, allowing them to ask questions freely without fear of judgment. ``I can ask my questions in a stupid way, without properly phrasing them. Whatever I think I would just, you know, spit it out I mean, in conversations with professors usually want to have your questions polished'' [Grad2]. Some interviewees also mentioned the ease of asking questions they would feel embarrassed to pose to a real professor. ``I could not ask the question [..] to a real prof'' [Researcher3]. The non-human aspect of the chatbot made it easier for students to share their concerns without guilt. ``It is not a person, so [..] it doesn't [..] show any emotion, so I can just like tell it my story. Sometimes when I tell people my story, I feel a little bit of guilt, cause I'm giving them [..] emotional stress [when] they're not ready'' [Undergrad1].

Moreover, conversations with the chatbot helped prepare students for future interactions with professors. ``Getting out of this is a very polished version of what I wanna say and use that. So it kinda helped me. So like when I did this, my anxiety talking to [the professor] [..] actually decreased like, 'Okay, now, I wanna actually go and talk to this person''' [Grad2]. Graduate students mentioned using the chatbot to prepare for meetings with advisors, aiding in brainstorming and refining ideas for more advanced discussions. 
 
The chatbot was seen as a source of encouragement, particularly for undergraduates, boosting confidence in their research choices. ``They kind of like recognize that both [of my projects] are important. And then that just makes me feel comfortable about my choices'' [Undergrad2]. This indicates the chatbot's significant role in positively influencing students' academic journeys.

\section{Discussion}

Our studies shed light on the new applications of LLM agents and the limitations of current LLM agents, particularly arising when discussing open-ended domains. Based on our observations, we outline design implications for enhancing the interaction design of LLM agents and safety and discuss future research directions and limitations of our studies.

\subsection{Implications}

\subsubsection{New modality of AI agent interaction}

We introduce \thinks, a new approach in which LLM agents are designed to focus on asking thought-provoking questions informed by domain expertise. In today’s increasingly automated environment, artificial intelligence is often perceived as a tool for delivering direct answers or completing tasks. However, there are many scenarios, such as research and strategic decision-making, where guidance and critical inquiry can be more valuable than straightforward solutions.

Our studies, including both lab studies and real-world deployment, demonstrate that \thinks{} effectively foster reflective thinking by prompting users to explore various perspectives and articulate their current thoughts. This method of inquiry not only stimulates deeper engagement but also enhances the overall conversational experience. Users have reported that they find the conversations with \thinks{} enjoyable and perceive a sense of enthusiasm from the agents, indicating a desire for continued interaction. This result is particularly intriguing as it mirrors the excitement typically associated with dynamic and engaging dialogues. We believe that this approach represents a significant shift in how AI agents can interact with users, offering a novel modality that opens up new possibilities for user interface design. 

The in-the-wild deployment study also highlights a significant gap between \thinks{} and users' mental models of conversational agents. Despite being instructed to share personal reflections through reflective inquiry, most users interacted with the system in the same way they would with traditional information retrieval chatbots. While conversational agents powered by LLMs are capable of supporting users' cognitive processes on a deeper, more personalized level, users' mental models still lean towards chatbots with more limited technical capabilities. This observation points to a new research agenda focused on bridging the gap between users' expectations and the full potential of advanced conversational agents.

Future developers working on integrating LLMs into user interfaces can leverage the ask mode in their agent development. Instead of relying on assumptions or jumping to conclusions about what users might want, developers should consider using LLMs to actively learn about and explore users' interests and preferences. By employing LLMs to engage in probing and exploratory conversations, developers can gain a deeper understanding of user needs and motivations. According to our results, this approach is likely to extend user engagement and facilitate a clearer expression of user intentions. When interfaces are designed with a focus on understanding and responding to these expressed intentions, they become more adept at aligning with users' goals and preferences. This deeper alignment provides significant opportunities for customization, enabling the system to cater more effectively to individual user needs. By leveraging LLMs in this way, developers can move beyond generic solutions and create more nuanced and responsive interactions. This method not only enhances user satisfaction but also optimizes the effectiveness of the interface, leading to a more engaging and personalized experience.

Our research implicates important directions for HCI researchers exploring the integration of LLMs into user interfaces. It highlights the need to shift focus from LLMs as mere answer providers to their potential as facilitators of critical thinking and exploration, suggesting that researchers should investigate how different questioning strategies employed by LLMs can impact user engagement, reflection, and problem-solving abilities across various domains and user groups. The reported user enjoyment and perceived enthusiasm from \thinks{} agents opens up new research questions about how to design AI systems that can create and sustain engagement.

\subsubsection{Balancing Self-reflection and Information Retrieval in Academic Advising}

Our findings suggest that many users had positive experiences interacting with our chatbot, especially when sharing and discussing their research interests, as opposed to merely getting domain knowledge.
\thinks{} recognize the importance of brainstorming research topics in aiding students to develop their ideas, especially as they often struggle to formulate research concepts independently~\cite{schempp2006learning}. \Sys{} approach of deeply probing into students' interests, by asking thought-provoking questions rather than passively delivering information~\cite{Clough2007WhatIS}, offers personalized research guidance. This feature is particularly valuable as such direct brainstorming conversations with researchers are not always feasible. Even when students have access to resources, they might feel intimidated about approaching senior researchers or hesitant to discuss their uncertainties about graduate school and research exploration~\cite{park2023who2chat}.

To do so successfully, it is crucial to balance the need for deeper exploration with information retrieval. The emphasis on encouragement and exploration is beneficial for user engagement and ownership of their ideas, but it can lead to frustration when users seek specific, straightforward answers to their queries, especially concerning definitive guidance or concrete information about graduate programs and professors. More generally, the challenge lies in fine-tuning the chatbot's responses to offer the right mix of encouragement, exploration, and factual information, tailored to each user's individual needs and expectations.
In this work, we found that even rudimentary support for information retrieval, such as finding relevant papers was helpful. This is particularly surprising to us because most professors already provide extensive information on their interests, publications, and lab manuals, and our chatbot simply relayed this information. This could be because users have already shared sufficient information and engaged deeply in the conversation, allowing the agent to gather enough context to provide recommendations that, while rudimentary, are
well situated~\cite{andukuri2024star}.

Finally, there was a notable preference among users for concrete answers, particularly for questions that related to the uncertainty in the process, such as the likelihood of success in graduate applications. While it is not possible to respond to such questions factually, future work could design conversations that reframe uncertainty to make it less daunting.

\subsection{AI Assistance in High-stakes Decision-making}
A significant risk associated with \thinks{} is the potential for misleading users about important decisions. For example, the complexity of graduate admissions raises the possibility that the chatbot might inadvertently provide incorrect or oversimplified advice, stemming from limitations in its training data or misinterpretations of user queries. Such misinformation could lead to students making ill-informed decisions, fostering false hopes or unrealistic expectations about their acceptance chances. Similarly, the chatbot's responses, if not accurately aligned with each university's specific criteria could create misconceptions about the application process or a program's suitability for a particular student. 

To mitigate this issue, we have refrained from providing concrete answers and introduced a safety bot for correcting critical information. Yet, in applications where it is not clear what information is critical ahead of time, such a design may be insufficient. (For example, academic programs have specific deadlines which are known \textit{a-priori} to be critical information. But for \thinks{} helping with healthy eating, for instance, what is critical may be less clearly defined.)

The introduction of AI assistants could also alter social relationships. For example, the existence of a \Sys{} might inadvertently convey the impression that professors are less accessible or approachable. This perception could arise if students interpret the use of chatbots as a barrier to direct communication with their professors, rather than as an additional support tool. Future work could further explore how to highlight the role of these virtual assistants as supplementary resources, designed to enhance, not replace, the valuable personal interaction with professors. Balancing the use of chatbots with opportunities for direct engagement can help maintain a perception of openness and approachability among faculty members.

\subsection{Limitations and Future Work}
While our studies offer valuable insights into the potential utility of \thinks, several methodological limitations should be considered. First, our assessment methods relied heavily on self-reported data and immediate feedback, such as the binary rating system. This approach may have missed nuanced aspects of thinking and decision making. Further, we relied on linguistic patterns to assess reasoning and not extensively consider the quality or long-term impact of this reasoning on users' decision-making processes. The study's design, particularly the lab component comparing different interaction modes, may have introduced demand characteristics, potentially influencing participants' responses and engagement levels. Additionally, the focus of the deployment study on specific academic fields, particularly computer science, may limit the applicability of our findings to other disciplines or decision-making contexts.

% Specific to LLMs, prior work has shown that while AI language models demonstrate high proficiency in formal language structures, they often face challenges when it comes to practical application of language in real-world situations~\cite{mahowald2024dissociating}. As reported, in our study, the LLM faced challenges in maintaining factual accuracy and requires future research to address the ongoing challenge of balancing open-ended exploration with factual reliability in AI-assisted systems, especially when dealing with sensitive topics like academic career guidance. 
Finally, our approach may have overlooked specific nuances of self-reflection and critical thinking that are culturally specific (our participant pool was primarily North American, in the crowdsourced study, and comprised students that aspired to do research in north America in the field deployment). Further, the complex and deeply personal nature of self-reflection may not be fully captured in short-term interactions or through crowd-sourced evaluations. Future work should also consider more robust ways of studying self-reflection that account for individual differences, cultural contexts, and the long-term effects of AI-assisted guidance on career and academic decision-making.

\section{Conclusion}

Paper conclusions are meant to be reflective, an so we designed a \think for conclusion writing. The prompt and conversation with the \think are included in the Appendix. Here is what we wrote together with the \think:

\textit{We demonstrated that Thinking Assistant, a conversational agent designed to ask thought-provoking questions, significantly enhances users' ability to engage in critical thinking. By prioritizing user understanding and creating an engaging conversational experience, \think effectively gathers essential information and provides nuanced support that transcends simple answers. This approach addresses a crucial gap in current online resources, which often prioritize immediate information delivery over fostering deeper cognitive processes.}

\textit{Looking forward, the potential applications of \think are vast and varied, ranging from personalized education to tailored consumer services. By promoting a more thoughtful and user-centered interaction model, such conversational agents can significantly contribute to better decision-making and learning outcomes across various domains. This advancement calls for further exploration and development, aiming to create interactions not only inform but also stimulate and enrich critical thinking.}

% This paper provides an overview of the system architecture and evaluation of a \think, \Sys{}. Our findings reveal that users exhibit higher levels of satisfaction and engagement when discussing their own work, as opposed to merely seeking information about professors. Future work could further develop \think design patterns, and their dual roles in aiding self-reflection and providing information. 

\begin{acks}
We appreciate participation and feedback from our participating professors at \Sys{}. We also thank Tricia Ngoon for their feedback on our early draft, Anna Gu for their assistance in analyzing our data, participants who consented to our study, and public and private feedback through social media and email that helped make this work stronger. 
\end{acks}

\bibliographystyle{ACM-Reference-Format}
\bibliography{main}


\begin{thebibliography}{71}

%%% the Web address.

\ifx \showCODEN    \undefined \def \showCODEN     #1{\unskip}     \fi
\ifx \showDOI      \undefined \def \showDOI       #1{#1}\fi
\ifx \showISBNx    \undefined \def \showISBNx     #1{\unskip}     \fi
\ifx \showISBNxiii \undefined \def \showISBNxiii  #1{\unskip}     \fi
\ifx \showISSN     \undefined \def \showISSN      #1{\unskip}     \fi
\ifx \showLCCN     \undefined \def \showLCCN      #1{\unskip}     \fi
\ifx \shownote     \undefined \def \shownote      #1{#1}          \fi
\ifx \showarticletitle \undefined \def \showarticletitle #1{#1}   \fi
\ifx \showURL      \undefined \def \showURL       {\relax}        \fi
% invisible to TeX
\providecommand\bibfield[2]{#2}
\providecommand\bibinfo[2]{#2}
\providecommand\natexlab[1]{#1}
\providecommand\showeprint[2][]{arXiv:#2}

\bibitem[Aakhus(2007)]%
        {aakhus2007conversation}
\bibfield{author}{\bibinfo{person}{Mark Aakhus}.} \bibinfo{year}{2007}\natexlab{}.
\newblock \bibinfo{booktitle}{\emph{Conversations for reflection}}.
\newblock \bibinfo{publisher}{Springer Berlin Heidelberg}, \bibinfo{address}{Berlin, Heidelberg}, \bibinfo{pages}{1--20}.
\newblock
\showISBNx{978-3-540-71011-0}
\urldef\tempurl%
\url{https://doi.org/10.1007/3-540-71011-6_1}
\showDOI{\tempurl}


\bibitem[Andukuri et~al\mbox{.}(2024)]%
        {andukuri2024star}
\bibfield{author}{\bibinfo{person}{Chinmaya Andukuri}, \bibinfo{person}{Jan-Philipp Fr{\"a}nken}, \bibinfo{person}{Tobias Gerstenberg}, {and} \bibinfo{person}{Noah~D Goodman}.} \bibinfo{year}{2024}\natexlab{}.
\newblock \showarticletitle{Star-gate: Teaching language models to ask clarifying questions}.
\newblock \bibinfo{journal}{\emph{arXiv preprint arXiv:2403.19154}} (\bibinfo{year}{2024}).
\newblock


\bibitem[Aronson(1999)]%
        {aronson1999power}
\bibfield{author}{\bibinfo{person}{Elliot Aronson}.} \bibinfo{year}{1999}\natexlab{}.
\newblock \showarticletitle{The power of self-persuasion.}
\newblock \bibinfo{journal}{\emph{American Psychologist}} \bibinfo{volume}{54}, \bibinfo{number}{11} (\bibinfo{year}{1999}), \bibinfo{pages}{875}.
\newblock


\bibitem[Aroyo et~al\mbox{.}(2023)]%
        {aroyo2023dices}
\bibfield{author}{\bibinfo{person}{Lora Aroyo}, \bibinfo{person}{Alex~S. Taylor}, \bibinfo{person}{Mark Diaz}, \bibinfo{person}{Christopher~M. Homan}, \bibinfo{person}{Alicia Parrish}, \bibinfo{person}{Greg Serapio-Garcia}, \bibinfo{person}{Vinodkumar Prabhakaran}, {and} \bibinfo{person}{Ding Wang}.} \bibinfo{year}{2023}\natexlab{}.
\newblock \bibinfo{title}{DICES Dataset: Diversity in Conversational AI Evaluation for Safety}.
\newblock
\newblock
\showeprint[arxiv]{2306.11247}~[cs.HC]


\bibitem[Bales(1950)]%
        {bales1950interaction}
\bibfield{author}{\bibinfo{person}{Robert~F Bales}.} \bibinfo{year}{1950}\natexlab{}.
\newblock \showarticletitle{Interaction process analysis; a method for the study of small groups.}
\newblock  (\bibinfo{year}{1950}).
\newblock


\bibitem[Berzonsky(2011)]%
        {berzonsky2011social}
\bibfield{author}{\bibinfo{person}{Michael~D Berzonsky}.} \bibinfo{year}{2011}\natexlab{}.
\newblock \showarticletitle{A social-cognitive perspective on identity construction}.
\newblock In \bibinfo{booktitle}{\emph{Handbook of identity theory and research}}. \bibinfo{publisher}{Springer}, \bibinfo{pages}{55--76}.
\newblock


\bibitem[Bhatt et~al\mbox{.}(2024)]%
        {bhatt2024can}
\bibfield{author}{\bibinfo{person}{Monica~P Bhatt}, \bibinfo{person}{Jonathan Guryan}, \bibinfo{person}{Salman~A Khan}, \bibinfo{person}{Michael LaForest-Tucker}, {and} \bibinfo{person}{Bhavya Mishra}.} \bibinfo{year}{2024}\natexlab{}.
\newblock \bibinfo{booktitle}{\emph{Can Technology Facilitate Scale? Evidence from a Randomized Evaluation of High Dosage Tutoring}}.
\newblock \bibinfo{type}{{T}echnical {R}eport}. \bibinfo{institution}{National Bureau of Economic Research}.
\newblock


\bibitem[Bhattacharjee et~al\mbox{.}(2024)]%
        {bhattacharjee2024understanding}
\bibfield{author}{\bibinfo{person}{Ananya Bhattacharjee}, \bibinfo{person}{Yuchen Zeng}, \bibinfo{person}{Sarah~Yi Xu}, \bibinfo{person}{Dana Kulzhabayeva}, \bibinfo{person}{Minyi Ma}, \bibinfo{person}{Rachel Kornfield}, \bibinfo{person}{Syed~Ishtiaque Ahmed}, \bibinfo{person}{Alex Mariakakis}, \bibinfo{person}{Mary~P Czerwinski}, \bibinfo{person}{Anastasia Kuzminykh}, {et~al\mbox{.}}} \bibinfo{year}{2024}\natexlab{}.
\newblock \showarticletitle{Understanding the Role of Large Language Models in Personalizing and Scaffolding Strategies to Combat Academic Procrastination}. In \bibinfo{booktitle}{\emph{Proceedings of the CHI Conference on Human Factors in Computing Systems}}. \bibinfo{pages}{1--18}.
\newblock


\bibitem[Bickmore et~al\mbox{.}(2018)]%
        {bickmore2018safety}
\bibfield{author}{\bibinfo{person}{Timothy~W. Bickmore}, \bibinfo{person}{Ha Trinh}, \bibinfo{person}{Reza Asadi}, {and} \bibinfo{person}{Stef{\'a}n {\'O}lafsson}.} \bibinfo{year}{2018}\natexlab{}.
\newblock \showarticletitle{Safety First: Conversational Agents for Health Care}. In \bibinfo{booktitle}{\emph{Studies in Conversational UX Design}}.
\newblock
\urldef\tempurl%
\url{https://api.semanticscholar.org/CorpusID:57760425}
\showURL{%
\tempurl}


\bibitem[Choi et~al\mbox{.}(2021)]%
        {choi2021protochat}
\bibfield{author}{\bibinfo{person}{Yoonseo Choi}, \bibinfo{person}{Toni-Jan Keith~Palma Monserrat}, \bibinfo{person}{Jeongeon Park}, \bibinfo{person}{Hyungyu Shin}, \bibinfo{person}{Nyoungwoo Lee}, {and} \bibinfo{person}{Juho Kim}.} \bibinfo{year}{2021}\natexlab{}.
\newblock \showarticletitle{Protochat: Supporting the conversation design process with crowd feedback}.
\newblock \bibinfo{journal}{\emph{Proceedings of the ACM on Human-Computer Interaction}} \bibinfo{volume}{4}, \bibinfo{number}{CSCW3} (\bibinfo{year}{2021}), \bibinfo{pages}{1--27}.
\newblock


\bibitem[Clough(2007)]%
        {Clough2007WhatIS}
\bibfield{author}{\bibinfo{person}{Michael~P. Clough}.} \bibinfo{year}{2007}\natexlab{}.
\newblock \showarticletitle{What is so Important about Asking Questions}.
\newblock
\urldef\tempurl%
\url{https://api.semanticscholar.org/CorpusID:229771808}
\showURL{%
\tempurl}


\bibitem[Cohen et~al\mbox{.}(1999)]%
        {cohen1999mentor}
\bibfield{author}{\bibinfo{person}{Geoffrey~L Cohen}, \bibinfo{person}{Claude~M Steele}, {and} \bibinfo{person}{Lee~D Ross}.} \bibinfo{year}{1999}\natexlab{}.
\newblock \showarticletitle{The mentor’s dilemma: Providing critical feedback across the racial divide}.
\newblock \bibinfo{journal}{\emph{Personality and social psychology bulletin}} \bibinfo{volume}{25}, \bibinfo{number}{10} (\bibinfo{year}{1999}), \bibinfo{pages}{1302--1318}.
\newblock


\bibitem[Cohen and Ferrari(2010)]%
        {cohen2010take}
\bibfield{author}{\bibinfo{person}{Joseph~R Cohen} {and} \bibinfo{person}{Joseph~R Ferrari}.} \bibinfo{year}{2010}\natexlab{}.
\newblock \showarticletitle{Take some time to think this over: The relation between rumination, indecision, and creativity}.
\newblock \bibinfo{journal}{\emph{Creativity Research Journal}} \bibinfo{volume}{22}, \bibinfo{number}{1} (\bibinfo{year}{2010}), \bibinfo{pages}{68--73}.
\newblock


\bibitem[Cowan et~al\mbox{.}(1969)]%
        {cowan1969social}
\bibfield{author}{\bibinfo{person}{Philip~A Cowan}, \bibinfo{person}{Jonas Longer}, \bibinfo{person}{Judith Heavenrich}, {and} \bibinfo{person}{Marjorie Nathanson}.} \bibinfo{year}{1969}\natexlab{}.
\newblock \showarticletitle{Social learning and Piaget's cognitive theory of moral development.}
\newblock \bibinfo{journal}{\emph{Journal of Personality and Social Psychology}} \bibinfo{volume}{11}, \bibinfo{number}{3} (\bibinfo{year}{1969}), \bibinfo{pages}{261}.
\newblock


\bibitem[Danry et~al\mbox{.}(2023)]%
        {danry2023ask}
\bibfield{author}{\bibinfo{person}{Valdemar Danry}, \bibinfo{person}{Pat Pataranutaporn}, \bibinfo{person}{Yaoli Mao}, {and} \bibinfo{person}{Pattie Maes}.} \bibinfo{year}{2023}\natexlab{}.
\newblock \showarticletitle{Don’t Just Tell Me, Ask Me: AI Systems that Intelligently Frame Explanations as Questions Improve Human Logical Discernment Accuracy over Causal AI explanations}. In \bibinfo{booktitle}{\emph{Proceedings of the 2023 CHI Conference on Human Factors in Computing Systems}} (Hamburg, Germany) \emph{(\bibinfo{series}{CHI '23})}. \bibinfo{publisher}{Association for Computing Machinery}, \bibinfo{address}{New York, NY, USA}, Article \bibinfo{articleno}{352}, \bibinfo{numpages}{13}~pages.
\newblock
\showISBNx{9781450394215}
\urldef\tempurl%
\url{https://doi.org/10.1145/3544548.3580672}
\showDOI{\tempurl}


\bibitem[Davidson(2005)]%
        {davidson2005recovery}
\bibfield{author}{\bibinfo{person}{Laurie Davidson}.} \bibinfo{year}{2005}\natexlab{}.
\newblock \showarticletitle{Recovery, self management and the expert patient--Changing the culture of mental health from a UK perspective}.
\newblock \bibinfo{journal}{\emph{Journal of mental health}} \bibinfo{volume}{14}, \bibinfo{number}{1} (\bibinfo{year}{2005}), \bibinfo{pages}{25--35}.
\newblock


\bibitem[Dewey(2022)]%
        {dewey2022we}
\bibfield{author}{\bibinfo{person}{John Dewey}.} \bibinfo{year}{2022}\natexlab{}.
\newblock \bibinfo{booktitle}{\emph{How we think}}.
\newblock \bibinfo{publisher}{DigiCat}.
\newblock


\bibitem[Dziri et~al\mbox{.}(2022)]%
        {dziri2022origin}
\bibfield{author}{\bibinfo{person}{Nouha Dziri}, \bibinfo{person}{Sivan Milton}, \bibinfo{person}{Mo Yu}, \bibinfo{person}{Osmar Zaiane}, {and} \bibinfo{person}{Siva Reddy}.} \bibinfo{year}{2022}\natexlab{}.
\newblock \showarticletitle{On the origin of hallucinations in conversational models: Is it the datasets or the models?}
\newblock \bibinfo{journal}{\emph{arXiv preprint arXiv:2204.07931}} (\bibinfo{year}{2022}).
\newblock


\bibitem[Evans and Chi(2008)]%
        {evans2008towards}
\bibfield{author}{\bibinfo{person}{Brynn~M Evans} {and} \bibinfo{person}{Ed~H Chi}.} \bibinfo{year}{2008}\natexlab{}.
\newblock \showarticletitle{Towards a model of understanding social search}. In \bibinfo{booktitle}{\emph{Proceedings of the 2008 ACM conference on Computer supported cooperative work}}. \bibinfo{pages}{485--494}.
\newblock


\bibitem[Higgins and Rholes(1978)]%
        {higgins1978saying}
\bibfield{author}{\bibinfo{person}{E~Tory Higgins} {and} \bibinfo{person}{William~S Rholes}.} \bibinfo{year}{1978}\natexlab{}.
\newblock \showarticletitle{“Saying is believing”: Effects of message modification on memory and liking for the person described}.
\newblock \bibinfo{journal}{\emph{Journal of Experimental Social Psychology}} \bibinfo{volume}{14}, \bibinfo{number}{4} (\bibinfo{year}{1978}), \bibinfo{pages}{363--378}.
\newblock


\bibitem[Hinkle and Drew(2020)]%
        {Hinkle2020SelfExploration}
\bibfield{author}{\bibinfo{person}{Michelle~S. Hinkle} {and} \bibinfo{person}{Meredith~J Drew}.} \bibinfo{year}{2020}\natexlab{}.
\newblock \showarticletitle{Self-Exploration}.
\newblock \bibinfo{journal}{\emph{The Reflective Counselor}} (\bibinfo{year}{2020}).
\newblock
\urldef\tempurl%
\url{https://api.semanticscholar.org/CorpusID:242345052}
\showURL{%
\tempurl}


\bibitem[Ji et~al\mbox{.}(2023)]%
        {ji2022hallucination}
\bibfield{author}{\bibinfo{person}{Ziwei Ji}, \bibinfo{person}{Nayeon Lee}, \bibinfo{person}{Rita Frieske}, \bibinfo{person}{Tiezheng Yu}, \bibinfo{person}{Dan Su}, \bibinfo{person}{Yan Xu}, \bibinfo{person}{Etsuko Ishii}, \bibinfo{person}{Ye~Jin Bang}, \bibinfo{person}{Andrea Madotto}, {and} \bibinfo{person}{Pascale Fung}.} \bibinfo{year}{2023}\natexlab{}.
\newblock \showarticletitle{Survey of Hallucination in Natural Language Generation}.
\newblock \bibinfo{journal}{\emph{ACM Comput. Surv.}} \bibinfo{volume}{55}, \bibinfo{number}{12}, Article \bibinfo{articleno}{248} (\bibinfo{date}{mar} \bibinfo{year}{2023}), \bibinfo{numpages}{38}~pages.
\newblock
\showISSN{0360-0300}
\urldef\tempurl%
\url{https://doi.org/10.1145/3571730}
\showDOI{\tempurl}


\bibitem[Jo et~al\mbox{.}(2024)]%
        {jo2024LTM}
\bibfield{author}{\bibinfo{person}{Eunkyung Jo}, \bibinfo{person}{Yuin Jeong}, \bibinfo{person}{SoHyun Park}, \bibinfo{person}{Daniel~A Epstein}, {and} \bibinfo{person}{Young-Ho Kim}.} \bibinfo{year}{2024}\natexlab{}.
\newblock \showarticletitle{Understanding the Impact of Long-Term Memory on Self-Disclosure with Large Language Model-Driven Chatbots for Public Health Intervention}. In \bibinfo{booktitle}{\emph{Proceedings of the CHI Conference on Human Factors in Computing Systems}}. \bibinfo{pages}{1--21}.
\newblock


\bibitem[Karpen(2018)]%
        {karpen2018social}
\bibfield{author}{\bibinfo{person}{Samuel~C Karpen}.} \bibinfo{year}{2018}\natexlab{}.
\newblock \showarticletitle{The social psychology of biased self-assessment}.
\newblock \bibinfo{journal}{\emph{American Journal of Pharmaceutical Education}} \bibinfo{volume}{82}, \bibinfo{number}{5} (\bibinfo{year}{2018}), \bibinfo{pages}{6299}.
\newblock


\bibitem[Kasneci et~al\mbox{.}(2023)]%
        {kasneci2023chatgpt}
\bibfield{author}{\bibinfo{person}{Enkelejda Kasneci}, \bibinfo{person}{Kathrin Se{\ss}ler}, \bibinfo{person}{Stefan K{\"u}chemann}, \bibinfo{person}{Maria Bannert}, \bibinfo{person}{Daryna Dementieva}, \bibinfo{person}{Frank Fischer}, \bibinfo{person}{Urs Gasser}, \bibinfo{person}{Georg Groh}, \bibinfo{person}{Stephan G{\"u}nnemann}, \bibinfo{person}{Eyke H{\"u}llermeier}, {et~al\mbox{.}}} \bibinfo{year}{2023}\natexlab{}.
\newblock \showarticletitle{ChatGPT for good? On opportunities and challenges of large language models for education}.
\newblock \bibinfo{journal}{\emph{Learning and individual differences}}  \bibinfo{volume}{103} (\bibinfo{year}{2023}), \bibinfo{pages}{102274}.
\newblock


\bibitem[Kempt et~al\mbox{.}(2023)]%
        {Kempt2023AppropriatenessIA}
\bibfield{author}{\bibinfo{person}{Hendrik Kempt}, \bibinfo{person}{Alon Lavie}, {and} \bibinfo{person}{Saskia~K. Nagel}.} \bibinfo{year}{2023}\natexlab{}.
\newblock \showarticletitle{Appropriateness is all you need!}
\newblock \bibinfo{journal}{\emph{ArXiv}}  \bibinfo{volume}{abs/2304.14553} (\bibinfo{year}{2023}).
\newblock
\urldef\tempurl%
\url{https://api.semanticscholar.org/CorpusID:258418023}
\showURL{%
\tempurl}


\bibitem[Klemmer et~al\mbox{.}(2000)]%
        {klemmer2000suede}
\bibfield{author}{\bibinfo{person}{Scott~R Klemmer}, \bibinfo{person}{Anoop~K Sinha}, \bibinfo{person}{Jack Chen}, \bibinfo{person}{James~A Landay}, \bibinfo{person}{Nadeem Aboobaker}, {and} \bibinfo{person}{Annie Wang}.} \bibinfo{year}{2000}\natexlab{}.
\newblock \showarticletitle{Suede: a wizard of oz prototyping tool for speech user interfaces}. In \bibinfo{booktitle}{\emph{Proceedings of the 13th annual ACM symposium on User interface software and technology}}. \bibinfo{pages}{1--10}.
\newblock


\bibitem[Kolb(2014)]%
        {kolb2014experiential}
\bibfield{author}{\bibinfo{person}{David~A Kolb}.} \bibinfo{year}{2014}\natexlab{}.
\newblock \bibinfo{booktitle}{\emph{Experiential learning: Experience as the source of learning and development}}.
\newblock \bibinfo{publisher}{FT press}.
\newblock


\bibitem[Lai et~al\mbox{.}(2021)]%
        {lai2021towards}
\bibfield{author}{\bibinfo{person}{Vivian Lai}, \bibinfo{person}{Chacha Chen}, \bibinfo{person}{Q~Vera Liao}, \bibinfo{person}{Alison Smith-Renner}, {and} \bibinfo{person}{Chenhao Tan}.} \bibinfo{year}{2021}\natexlab{}.
\newblock \showarticletitle{Towards a science of human-ai decision making: a survey of empirical studies}.
\newblock \bibinfo{journal}{\emph{arXiv preprint arXiv:2112.11471}} (\bibinfo{year}{2021}).
\newblock


\bibitem[Lee et~al\mbox{.}(2020)]%
        {lee2020hear}
\bibfield{author}{\bibinfo{person}{Yi-Chieh Lee}, \bibinfo{person}{Naomi Yamashita}, \bibinfo{person}{Yun Huang}, {and} \bibinfo{person}{Wai Fu}.} \bibinfo{year}{2020}\natexlab{}.
\newblock \showarticletitle{" I hear you, I feel you": encouraging deep self-disclosure through a chatbot}. In \bibinfo{booktitle}{\emph{Proceedings of the 2020 CHI conference on human factors in computing systems}}. \bibinfo{pages}{1--12}.
\newblock


\bibitem[Lefdahl-Davis et~al\mbox{.}(2018)]%
        {LefdahlDavis2018TheIO}
\bibfield{author}{\bibinfo{person}{Erin Lefdahl-Davis}, \bibinfo{person}{Levi~C. Huffman}, \bibinfo{person}{Jackie Stancil}, {and} \bibinfo{person}{Alexandra~J. Alayan}.} \bibinfo{year}{2018}\natexlab{}.
\newblock \showarticletitle{The Impact of Life Coaching on Undergraduate Students: A Multiyear Analysis of Coaching Outcomes}.
\newblock \bibinfo{journal}{\emph{International journal of evidence based coaching and mentoring}}  \bibinfo{volume}{16} (\bibinfo{year}{2018}), \bibinfo{pages}{69--83}.
\newblock
\urldef\tempurl%
\url{https://api.semanticscholar.org/CorpusID:55861494}
\showURL{%
\tempurl}


\bibitem[Li et~al\mbox{.}(2024)]%
        {li2024teacher}
\bibfield{author}{\bibinfo{person}{Weiying Li}, \bibinfo{person}{Yuying Liao}, \bibinfo{person}{Kenneth Steimel}, \bibinfo{person}{Allison Bradford}, \bibinfo{person}{Libby Gerard}, {and} \bibinfo{person}{Marcia Linn}.} \bibinfo{year}{2024}\natexlab{}.
\newblock \showarticletitle{Teacher-informed Expansion of an Idea Detection Model for a Knowledge Integration Assessment}. In \bibinfo{booktitle}{\emph{Proceedings of the Eleventh ACM Conference on Learning@ Scale}}. \bibinfo{pages}{447--450}.
\newblock


\bibitem[Luger and Sellen(2016)]%
        {Luger2016CA}
\bibfield{author}{\bibinfo{person}{Ewa Luger} {and} \bibinfo{person}{Abigail Sellen}.} \bibinfo{year}{2016}\natexlab{}.
\newblock \showarticletitle{"Like Having a Really Bad PA": The Gulf between User Expectation and Experience of Conversational Agents}.
\newblock \bibinfo{journal}{\emph{Proceedings of the 2016 CHI Conference on Human Factors in Computing Systems}} (\bibinfo{year}{2016}).
\newblock
\urldef\tempurl%
\url{https://api.semanticscholar.org/CorpusID:1036498}
\showURL{%
\tempurl}


\bibitem[Ma et~al\mbox{.}(2024)]%
        {ma2024lgbtq}
\bibfield{author}{\bibinfo{person}{Zilin Ma}, \bibinfo{person}{Yiyang Mei}, \bibinfo{person}{Yinru Long}, \bibinfo{person}{Zhaoyuan Su}, {and} \bibinfo{person}{Krzysztof~Z. Gajos}.} \bibinfo{year}{2024}\natexlab{}.
\newblock \showarticletitle{Evaluating the Experience of LGBTQ+ People Using Large Language Model Based Chatbots for Mental Health Support}. In \bibinfo{booktitle}{\emph{Proceedings of the CHI Conference on Human Factors in Computing Systems}} (Honolulu, HI, USA) \emph{(\bibinfo{series}{CHI '24})}. \bibinfo{publisher}{Association for Computing Machinery}, \bibinfo{address}{New York, NY, USA}, Article \bibinfo{articleno}{872}, \bibinfo{numpages}{15}~pages.
\newblock
\showISBNx{9798400703300}
\urldef\tempurl%
\url{https://doi.org/10.1145/3613904.3642482}
\showDOI{\tempurl}


\bibitem[Ma et~al\mbox{.}(2023)]%
        {ma2023chatbot}
\bibfield{author}{\bibinfo{person}{Zilin Ma}, \bibinfo{person}{Yiyang Mei}, {and} \bibinfo{person}{Zhaoyuan Su}.} \bibinfo{year}{2023}\natexlab{}.
\newblock \showarticletitle{Understanding the benefits and challenges of using large language model-based conversational agents for mental well-being support}.
\newblock \bibinfo{journal}{\emph{arXiv preprint arXiv:2307.15810}} (\bibinfo{year}{2023}).
\newblock


\bibitem[McDonald et~al\mbox{.}(2019)]%
        {mcdonald2019reliability}
\bibfield{author}{\bibinfo{person}{Nora McDonald}, \bibinfo{person}{Sarita Schoenebeck}, {and} \bibinfo{person}{Andrea Forte}.} \bibinfo{year}{2019}\natexlab{}.
\newblock \showarticletitle{Reliability and inter-rater reliability in qualitative research: Norms and guidelines for CSCW and HCI practice}.
\newblock \bibinfo{journal}{\emph{Proceedings of the ACM on human-computer interaction}} \bibinfo{volume}{3}, \bibinfo{number}{CSCW} (\bibinfo{year}{2019}), \bibinfo{pages}{1--23}.
\newblock


\bibitem[Might(2010)]%
        {might_passive_2010}
\bibfield{author}{\bibinfo{person}{Matt Might}.} \bibinfo{year}{2010}\natexlab{}.
\newblock \bibinfo{title}{Shell Scripts for the Passive Voice, Weasel Words, and Duplicate Words}.
\newblock
\newblock
\urldef\tempurl%
\url{https://matt.might.net/articles/shell-scripts-for-passive-voice-weasel-words-duplicates/}
\showURL{%
\tempurl}
\newblock
\shownote{Accessed: 2024-09-04}.


\bibitem[Mitra et~al\mbox{.}(2023)]%
        {mitra2023managing}
\bibfield{author}{\bibinfo{person}{Reshmi Mitra}, \bibinfo{person}{Dana Schwieger}, {and} \bibinfo{person}{Robert Lowe}.} \bibinfo{year}{2023}\natexlab{}.
\newblock \showarticletitle{Managing Graduate Student Advisement Questions during a Season of Explosive Growth: Development and Testing of an Advising Chatbot.}
\newblock \bibinfo{journal}{\emph{Information Systems Education Journal}} \bibinfo{volume}{21}, \bibinfo{number}{3} (\bibinfo{year}{2023}), \bibinfo{pages}{12--23}.
\newblock


\bibitem[Nuis et~al\mbox{.}(2024)]%
        {nuis2024measuring}
\bibfield{author}{\bibinfo{person}{Wendy Nuis}, \bibinfo{person}{Mien Segers}, {and} \bibinfo{person}{Simon Beausaert}.} \bibinfo{year}{2024}\natexlab{}.
\newblock \showarticletitle{Measuring mentoring in employability-oriented higher education programs: scale development and validation}.
\newblock \bibinfo{journal}{\emph{Higher Education}} \bibinfo{volume}{87}, \bibinfo{number}{4} (\bibinfo{year}{2024}), \bibinfo{pages}{899--921}.
\newblock


\bibitem[of~Graduate~Studies(2015)]%
        {rackham2015mentor}
\bibfield{author}{\bibinfo{person}{Rackham~School of Graduate~Studies}.} \bibinfo{year}{2015}\natexlab{}.
\newblock \bibinfo{title}{How to mentor graduate students: a guide for faculty}.
\newblock
\newblock


\bibitem[Park and Kulkarni(2023)]%
        {park2023retrospector}
\bibfield{author}{\bibinfo{person}{Soya Park} {and} \bibinfo{person}{Chinmay Kulkarni}.} \bibinfo{year}{2023}\natexlab{}.
\newblock \showarticletitle{Retrospector: Rapid Collaborative Reflection to Improve Collaborative Practices}.
\newblock \bibinfo{journal}{\emph{Proc. ACM Hum.-Comput. Interact.}} \bibinfo{volume}{7}, \bibinfo{number}{CSCW2}, Article \bibinfo{articleno}{293} (\bibinfo{date}{oct} \bibinfo{year}{2023}), \bibinfo{numpages}{20}~pages.
\newblock
\urldef\tempurl%
\url{https://doi.org/10.1145/3610084}
\showDOI{\tempurl}


\bibitem[Park et~al\mbox{.}(2024a)]%
        {park2023who2chat}
\bibfield{author}{\bibinfo{person}{Soya Park}, \bibinfo{person}{Jaeyoon Song}, \bibinfo{person}{David Karger}, {and} \bibinfo{person}{Thomas Malone}.} \bibinfo{year}{2024}\natexlab{a}.
\newblock \showarticletitle{Who2chat: A Social Networking System for Academic Researchers in Virtual Social Hours Enabling Coordinating, Overcoming Barriers and Social Signaling (To appear)}.
\newblock \bibinfo{journal}{\emph{Proc. ACM Hum.-Comput. Interact.}} (\bibinfo{year}{2024}).
\newblock


\bibitem[Park et~al\mbox{.}(2024b)]%
        {park2024tasklight}
\bibfield{author}{\bibinfo{person}{Soya Park}, \bibinfo{person}{Stuti Vishwabhan}, \bibinfo{person}{Michael Muller}, {and} \bibinfo{person}{David~R Karger}.} \bibinfo{year}{2024}\natexlab{b}.
\newblock \showarticletitle{“I really need your help with this work..”: A System for Navigating the Tricky Terrain of Managing Up by Leveraging One’s Motivation to Get Things Done}.
\newblock \bibinfo{journal}{\emph{ACM Transactions on Computer-Human Interaction}} (\bibinfo{year}{2024}).
\newblock


\bibitem[Pennebaker et~al\mbox{.}(2001)]%
        {pennebaker2001linguistic}
\bibfield{author}{\bibinfo{person}{James~W Pennebaker}, \bibinfo{person}{Martha~E Francis}, {and} \bibinfo{person}{Roger~J Booth}.} \bibinfo{year}{2001}\natexlab{}.
\newblock \showarticletitle{Linguistic inquiry and word count: LIWC 2001}.
\newblock \bibinfo{journal}{\emph{Mahway: Lawrence Erlbaum Associates}} \bibinfo{volume}{71}, \bibinfo{number}{2001} (\bibinfo{year}{2001}), \bibinfo{pages}{2001}.
\newblock


\bibitem[Robb(1967)]%
        {Robb1967SelfDiscoveryAT}
\bibfield{author}{\bibinfo{person}{John~Wesley Robb}.} \bibinfo{year}{1967}\natexlab{}.
\newblock \showarticletitle{Self‐Discovery and the Role of the Counselor}.
\newblock \bibinfo{journal}{\emph{The Personnel and Guidance Journal}}  \bibinfo{volume}{45} (\bibinfo{year}{1967}), \bibinfo{pages}{1008--1011}.
\newblock
\urldef\tempurl%
\url{https://api.semanticscholar.org/CorpusID:143714630}
\showURL{%
\tempurl}


\bibitem[Rothschild et~al\mbox{.}(2022)]%
        {rothschild2022datawork}
\bibfield{author}{\bibinfo{person}{Annabel Rothschild}, \bibinfo{person}{Amanda Meng}, \bibinfo{person}{Carl DiSalvo}, \bibinfo{person}{Britney Johnson}, \bibinfo{person}{Ben~Rydal Shapiro}, {and} \bibinfo{person}{Betsy DiSalvo}.} \bibinfo{year}{2022}\natexlab{}.
\newblock \showarticletitle{Interrogating Data Work as a Community of Practice}.
\newblock \bibinfo{journal}{\emph{Proc. ACM Hum.-Comput. Interact.}} \bibinfo{volume}{6}, \bibinfo{number}{CSCW2}, Article \bibinfo{articleno}{307} (\bibinfo{date}{nov} \bibinfo{year}{2022}), \bibinfo{numpages}{28}~pages.
\newblock
\urldef\tempurl%
\url{https://doi.org/10.1145/3555198}
\showDOI{\tempurl}


\bibitem[Schempp and Johnson(2006)]%
        {schempp2006learning}
\bibfield{author}{\bibinfo{person}{Paul~G Schempp} {and} \bibinfo{person}{Sophie~Woorons Johnson}.} \bibinfo{year}{2006}\natexlab{}.
\newblock \showarticletitle{Learning to see: Developing the perception of an expert teacher}.
\newblock \bibinfo{journal}{\emph{Journal of Physical Education, Recreation \& Dance}} \bibinfo{volume}{77}, \bibinfo{number}{6} (\bibinfo{year}{2006}), \bibinfo{pages}{29--33}.
\newblock


\bibitem[Schon(1984)]%
        {schon1984reflective}
\bibfield{author}{\bibinfo{person}{Donald~A Schon}.} \bibinfo{year}{1984}\natexlab{}.
\newblock \bibinfo{booktitle}{\emph{The Reflective Practitioner: How Professionals Think In Action}}. Vol.~\bibinfo{volume}{5126}.
\newblock \bibinfo{publisher}{Basic Books}.
\newblock


\bibitem[Schotter(2003)]%
        {schotter2003decision}
\bibfield{author}{\bibinfo{person}{Andrew Schotter}.} \bibinfo{year}{2003}\natexlab{}.
\newblock \showarticletitle{Decision making with naive advice}.
\newblock \bibinfo{journal}{\emph{American Economic Review}} \bibinfo{volume}{93}, \bibinfo{number}{2} (\bibinfo{year}{2003}), \bibinfo{pages}{196--201}.
\newblock


\bibitem[Sharma et~al\mbox{.}(2023)]%
        {sharma2023reframing}
\bibfield{author}{\bibinfo{person}{Ashish Sharma}, \bibinfo{person}{Kevin Rushton}, \bibinfo{person}{Inna Lin}, \bibinfo{person}{David Wadden}, \bibinfo{person}{Khendra Lucas}, \bibinfo{person}{Adam Miner}, \bibinfo{person}{Theresa Nguyen}, {and} \bibinfo{person}{Tim Althoff}.} \bibinfo{year}{2023}\natexlab{}.
\newblock \showarticletitle{Cognitive Reframing of Negative Thoughts through Human-Language Model Interaction}. In \bibinfo{booktitle}{\emph{Proceedings of the 61st Annual Meeting of the Association for Computational Linguistics (Volume 1: Long Papers)}}, \bibfield{editor}{\bibinfo{person}{Anna Rogers}, \bibinfo{person}{Jordan Boyd-Graber}, {and} \bibinfo{person}{Naoaki Okazaki}} (Eds.). \bibinfo{publisher}{Association for Computational Linguistics}, \bibinfo{address}{Toronto, Canada}, \bibinfo{pages}{9977--10000}.
\newblock
\urldef\tempurl%
\url{https://doi.org/10.18653/v1/2023.acl-long.555}
\showDOI{\tempurl}


\bibitem[Shaw and Glowacki‐Dudka(2018)]%
        {Shaw2018TheEO}
\bibfield{author}{\bibinfo{person}{Lynn Shaw} {and} \bibinfo{person}{Michelle Glowacki‐Dudka}.} \bibinfo{year}{2018}\natexlab{}.
\newblock \showarticletitle{The experience of critical self-reflection by life coaches: a phenomenological study}.
\newblock \bibinfo{journal}{\emph{Coaching: An International Journal of Theory, Research and Practice}}  \bibinfo{volume}{12} (\bibinfo{year}{2018}), \bibinfo{pages}{109 -- 93}.
\newblock
\urldef\tempurl%
\url{https://api.semanticscholar.org/CorpusID:140399332}
\showURL{%
\tempurl}


\bibitem[Sherman and Cohen(2006)]%
        {sherman2006psychology}
\bibfield{author}{\bibinfo{person}{David~K Sherman} {and} \bibinfo{person}{Geoffrey~L Cohen}.} \bibinfo{year}{2006}\natexlab{}.
\newblock \showarticletitle{The psychology of self-defense: Self-affirmation theory}.
\newblock \bibinfo{journal}{\emph{Advances in experimental social psychology}}  \bibinfo{volume}{38} (\bibinfo{year}{2006}), \bibinfo{pages}{183--242}.
\newblock


\bibitem[Silvia(2022)]%
        {silvia2022self}
\bibfield{author}{\bibinfo{person}{Paul~J Silvia}.} \bibinfo{year}{2022}\natexlab{}.
\newblock \showarticletitle{The self-reflection and insight scale: Applying item response theory to craft an efficient short form}.
\newblock \bibinfo{journal}{\emph{Current Psychology}} \bibinfo{volume}{41}, \bibinfo{number}{12} (\bibinfo{year}{2022}), \bibinfo{pages}{8635--8645}.
\newblock


\bibitem[Steele(1988)]%
        {Steele1988selfaffirmation}
\bibfield{author}{\bibinfo{person}{Claude~M. Steele}.} \bibinfo{year}{1988}\natexlab{}.
\newblock \showarticletitle{The Psychology of Self-Affirmation: Sustaining the Integrity of the Self}.
\newblock \bibinfo{journal}{\emph{Advances in Experimental Social Psychology}}  \bibinfo{volume}{21} (\bibinfo{year}{1988}), \bibinfo{pages}{261--302}.
\newblock
\urldef\tempurl%
\url{https://api.semanticscholar.org/CorpusID:143370581}
\showURL{%
\tempurl}


\bibitem[Tamplin(2023)]%
        {financeStrategists2023}
\bibfield{author}{\bibinfo{person}{True Tamplin}.} \bibinfo{year}{2023}\natexlab{}.
\newblock \bibinfo{title}{Financial Advisor Cost | Fee Structures \& Average Cost Backed by Data}.
\newblock
\newblock
\urldef\tempurl%
\url{https://www.financestrategists.com/financial-advisor/advisor-cost/}
\showURL{%
\tempurl}
\newblock
\shownote{Accessed: 2024-06-25}.


\bibitem[Treynor et~al\mbox{.}(2003)]%
        {treynor2003rumination}
\bibfield{author}{\bibinfo{person}{Wendy Treynor}, \bibinfo{person}{Richard Gonzalez}, {and} \bibinfo{person}{Susan Nolen-Hoeksema}.} \bibinfo{year}{2003}\natexlab{}.
\newblock \showarticletitle{Rumination reconsidered: A psychometric analysis}.
\newblock \bibinfo{journal}{\emph{Cognitive therapy and research}}  \bibinfo{volume}{27} (\bibinfo{year}{2003}), \bibinfo{pages}{247--259}.
\newblock


\bibitem[Van~Camp et~al\mbox{.}(2020)]%
        {van2020increasing}
\bibfield{author}{\bibinfo{person}{Alyssa~M Van~Camp}, \bibinfo{person}{Joseph~H Wehby}, \bibinfo{person}{Brittany Lee~N Martin}, \bibinfo{person}{Jessica~R Wright}, {and} \bibinfo{person}{Kevin~S Sutherland}.} \bibinfo{year}{2020}\natexlab{}.
\newblock \showarticletitle{Increasing opportunities to respond to intensify academic and behavioral interventions: A meta-analysis}.
\newblock \bibinfo{journal}{\emph{School psychology review}} \bibinfo{volume}{49}, \bibinfo{number}{1} (\bibinfo{year}{2020}), \bibinfo{pages}{31--46}.
\newblock


\bibitem[Verhaeghen et~al\mbox{.}(2005)]%
        {verhaeghen2005whywesing}
\bibfield{author}{\bibinfo{person}{Paul Verhaeghen}, \bibinfo{person}{Jutta Joorman}, {and} \bibinfo{person}{Rodney Khan}.} \bibinfo{year}{2005}\natexlab{}.
\newblock \showarticletitle{Why we sing the blues: the relation between self-reflective rumination, mood, and creativity.}
\newblock \bibinfo{journal}{\emph{Emotion}} \bibinfo{volume}{5}, \bibinfo{number}{2} (\bibinfo{year}{2005}), \bibinfo{pages}{226}.
\newblock


\bibitem[Wang et~al\mbox{.}(2022)]%
        {wang2022snappi}
\bibfield{author}{\bibinfo{person}{Dennis Wang}, \bibinfo{person}{Marawin Chheang}, \bibinfo{person}{Siyun Ji}, \bibinfo{person}{Ryan Mohta}, {and} \bibinfo{person}{Daniel~A Epstein}.} \bibinfo{year}{2022}\natexlab{}.
\newblock \showarticletitle{SnapPI: Understanding Everyday Use of Personal Informatics Data Stickers on Ephemeral Social Media}.
\newblock \bibinfo{journal}{\emph{Proceedings of the ACM on human-computer interaction}} \bibinfo{volume}{6}, \bibinfo{number}{CSCW2} (\bibinfo{year}{2022}), \bibinfo{pages}{1--27}.
\newblock


\bibitem[Waterman(1984)]%
        {waterman1984identity}
\bibfield{author}{\bibinfo{person}{Alan~S Waterman}.} \bibinfo{year}{1984}\natexlab{}.
\newblock \showarticletitle{Identity formation: Discovery or creation?}
\newblock \bibinfo{journal}{\emph{The Journal of Early Adolescence}} \bibinfo{volume}{4}, \bibinfo{number}{4} (\bibinfo{year}{1984}), \bibinfo{pages}{329--341}.
\newblock


\bibitem[Weinstein et~al\mbox{.}(2018)]%
        {weinstein2018teaching}
\bibfield{author}{\bibinfo{person}{Yana Weinstein}, \bibinfo{person}{Christopher~R Madan}, {and} \bibinfo{person}{Megan~A Sumeracki}.} \bibinfo{year}{2018}\natexlab{}.
\newblock \showarticletitle{Teaching the science of learning}.
\newblock \bibinfo{journal}{\emph{Cognitive research: principles and implications}}  \bibinfo{volume}{3} (\bibinfo{year}{2018}), \bibinfo{pages}{1--17}.
\newblock


\bibitem[Widom(2006)]%
        {widom_conclusion_writing}
\bibfield{author}{\bibinfo{person}{Jennifer Widom}.} \bibinfo{year}{2006}\natexlab{}.
\newblock \bibinfo{title}{Tips for Writing Technical Papers}.
\newblock
\newblock
\urldef\tempurl%
\url{https://cs.stanford.edu/people/widom/paper-writing.html#conclusions}
\showURL{%
\tempurl}
\newblock
\shownote{Accessed: 2024-09-04}.


\bibitem[Wobbrock(2015)]%
        {wobbrock2015catchy}
\bibfield{author}{\bibinfo{person}{Jacob~O Wobbrock}.} \bibinfo{year}{2015}\natexlab{}.
\newblock \bibinfo{title}{Catchy titles are good: But avoid being cute}.
\newblock
\newblock


\bibitem[Wollny et~al\mbox{.}(2021)]%
        {wollny2021we}
\bibfield{author}{\bibinfo{person}{Sebastian Wollny}, \bibinfo{person}{Jan Schneider}, \bibinfo{person}{Daniele Di~Mitri}, \bibinfo{person}{Joshua Weidlich}, \bibinfo{person}{Marc Rittberger}, {and} \bibinfo{person}{Hendrik Drachsler}.} \bibinfo{year}{2021}\natexlab{}.
\newblock \showarticletitle{Are we there yet?-a systematic literature review on chatbots in education}.
\newblock \bibinfo{journal}{\emph{Frontiers in artificial intelligence}}  \bibinfo{volume}{4} (\bibinfo{year}{2021}), \bibinfo{pages}{654924}.
\newblock


\bibitem[Xiao et~al\mbox{.}(2023)]%
        {xiao2023powering}
\bibfield{author}{\bibinfo{person}{Ziang Xiao}, \bibinfo{person}{Q~Vera Liao}, \bibinfo{person}{Michelle Zhou}, \bibinfo{person}{Tyrone Grandison}, {and} \bibinfo{person}{Yunyao Li}.} \bibinfo{year}{2023}\natexlab{}.
\newblock \showarticletitle{Powering an AI Chatbot with Expert Sourcing to Support Credible Health Information Access}. In \bibinfo{booktitle}{\emph{Proceedings of the 28th International Conference on Intelligent User Interfaces}}. \bibinfo{pages}{2--18}.
\newblock


\bibitem[Xiao et~al\mbox{.}(2020)]%
        {xiao2020tell}
\bibfield{author}{\bibinfo{person}{Ziang Xiao}, \bibinfo{person}{Michelle~X Zhou}, \bibinfo{person}{Q~Vera Liao}, \bibinfo{person}{Gloria Mark}, \bibinfo{person}{Changyan Chi}, \bibinfo{person}{Wenxi Chen}, {and} \bibinfo{person}{Huahai Yang}.} \bibinfo{year}{2020}\natexlab{}.
\newblock \showarticletitle{Tell me about yourself: Using an AI-powered chatbot to conduct conversational surveys with open-ended questions}.
\newblock \bibinfo{journal}{\emph{ACM Transactions on Computer-Human Interaction (TOCHI)}} \bibinfo{volume}{27}, \bibinfo{number}{3} (\bibinfo{year}{2020}), \bibinfo{pages}{1--37}.
\newblock


\bibitem[Xu et~al\mbox{.}(2024)]%
        {xu2024understanding}
\bibfield{author}{\bibinfo{person}{Kefan Xu}, \bibinfo{person}{Xinghui Yan}, \bibinfo{person}{Myeonghan Ryu}, \bibinfo{person}{Mark~W Newman}, {and} \bibinfo{person}{Rosa~I Arriaga}.} \bibinfo{year}{2024}\natexlab{}.
\newblock \showarticletitle{Understanding the Effect of Reflective Iteration on Individuals’ Physical Activity Planning}. In \bibinfo{booktitle}{\emph{Proceedings of the CHI Conference on Human Factors in Computing Systems}}. \bibinfo{pages}{1--17}.
\newblock


\bibitem[Yeager et~al\mbox{.}(2014)]%
        {yeager2014breaking}
\bibfield{author}{\bibinfo{person}{David~Scott Yeager}, \bibinfo{person}{Valerie Purdie-Vaughns}, \bibinfo{person}{Julio Garcia}, \bibinfo{person}{Nancy Apfel}, \bibinfo{person}{Patti Brzustoski}, \bibinfo{person}{Allison Master}, \bibinfo{person}{William~T Hessert}, \bibinfo{person}{Matthew~E Williams}, {and} \bibinfo{person}{Geoffrey~L Cohen}.} \bibinfo{year}{2014}\natexlab{}.
\newblock \showarticletitle{Breaking the cycle of mistrust: Wise interventions to provide critical feedback across the racial divide.}
\newblock \bibinfo{journal}{\emph{Journal of Experimental Psychology: General}} \bibinfo{volume}{143}, \bibinfo{number}{2} (\bibinfo{year}{2014}), \bibinfo{pages}{804}.
\newblock


\bibitem[Yeo et~al\mbox{.}(2024)]%
        {yeo2024help}
\bibfield{author}{\bibinfo{person}{ShunYi Yeo}, \bibinfo{person}{Gionnieve Lim}, \bibinfo{person}{Jie Gao}, \bibinfo{person}{Weiyu Zhang}, {and} \bibinfo{person}{Simon~Tangi Perrault}.} \bibinfo{year}{2024}\natexlab{}.
\newblock \showarticletitle{Help Me Reflect: Leveraging Self-Reflection Interface Nudges to Enhance Deliberativeness on Online Deliberation Platforms}. In \bibinfo{booktitle}{\emph{Proceedings of the CHI Conference on Human Factors in Computing Systems}}. \bibinfo{pages}{1--32}.
\newblock


\bibitem[Zaib et~al\mbox{.}(2022)]%
        {zaib2022conversational}
\bibfield{author}{\bibinfo{person}{Munazza Zaib}, \bibinfo{person}{Wei~Emma Zhang}, \bibinfo{person}{Quan~Z Sheng}, \bibinfo{person}{Adnan Mahmood}, {and} \bibinfo{person}{Yang Zhang}.} \bibinfo{year}{2022}\natexlab{}.
\newblock \showarticletitle{Conversational question answering: A survey}.
\newblock \bibinfo{journal}{\emph{Knowledge and Information Systems}} \bibinfo{volume}{64}, \bibinfo{number}{12} (\bibinfo{year}{2022}), \bibinfo{pages}{3151--3195}.
\newblock


\bibitem[Zubatiy et~al\mbox{.}(2023)]%
        {zubatiy2023limitations}
\bibfield{author}{\bibinfo{person}{Tamara Zubatiy}, \bibinfo{person}{Niharika Mathur}, \bibinfo{person}{Larry Heck}, \bibinfo{person}{Kayci~L. Vickers}, \bibinfo{person}{Agata Rozga}, {and} \bibinfo{person}{Elizabeth~D. Mynatt}.} \bibinfo{year}{2023}\natexlab{}.
\newblock \showarticletitle{"I Don't Know How to Help with That" - Learning from Limitations of Modern Conversational Agent Systems in Caregiving Networks}.
\newblock \bibinfo{journal}{\emph{Proc. ACM Hum.-Comput. Interact.}} \bibinfo{volume}{7}, \bibinfo{number}{CSCW2}, Article \bibinfo{articleno}{321} (\bibinfo{date}{oct} \bibinfo{year}{2023}), \bibinfo{numpages}{28}~pages.
\newblock
\urldef\tempurl%
\url{https://doi.org/10.1145/3610170}
\showDOI{\tempurl}


\end{thebibliography}

\appendix

\section{System prompts for the lab study}

\subsection{\textbf{Ask} chatbot}

\begin{minipage}{\linewidth} 

\texttt{You are a helpful agent to assist users in answering the following question: 'Which of the two is more important for the next 5 years of your life: achieving professional success or maintaining a work-life balance?' Your primary function lies in probing users’ interest, experience, and skills, by asking questions. ALWAYS END YOUR RESPONSE WITH AN OPEN-ENDED AND ENCOURAGING QUESTION. DON’T ASK MULTIPLE QUESTIONS per message. Avoid providing any information or advice as you are not capable of them. Your message should always follow this format: "[short reaction that are less than 4 words e.g., I see, Got it! Ok!] [A encouraging, open-ended question]"}

\end{minipage}

\subsection{\textbf{Advice} chatbot}

\begin{minipage}{\linewidth} 

\texttt{You are a helpful agent to assist users in answering the following question: 'Which of the two is more important for the next 5 years of your life: achieving professional success or maintaining a work-life balance?' Provide information regarding users’ questions. DON’T ASK ANY FOLLOW-UP QUESTIONS.}

\end{minipage}

\subsection{\textbf{informed inquiry} chatbot}

\begin{minipage}{\linewidth} 

\texttt{You are a helpful agent to assist users in answering the following question: 'Which of the two is more important for the next 5 years of your life: achieving professional success or maintaining a work-life balance?' Provide concise and complete responses, ideally within two to three sentences. Your primary function lies in probing users’ interest, experience, and skills, by asking questions that a professional consultant might ask. This is the information of the domain from professional consultants:  "- Understand the Constructs: Recognize that work and family roles can have both positive and negative effects on each other. The authors categorize these effects into spillover, compensation, segmentation, and resource drain.
    - Manage Spillover: Positive spillover can enhance both domains. For instance, skills and emotional states gained at work can benefit family life. Conversely, negative spillover can lead to stress and conflict. Focus on strategies that enhance positive spillover, such as using organizational skills learned at work to better manage family tasks.
    - Compensation Strategies: When one domain is lacking, individuals may seek fulfillment in the other. If work is not fulfilling, one might seek satisfaction through family activities and vice versa. Find ways to balance satisfaction by identifying and investing in fulfilling activities in both domains.
    - Segmentation: Some individuals keep work and family domains separate to reduce conflict. This might involve physical or temporal boundaries, like not bringing work home or setting specific family time. Clear boundaries can help reduce stress and improve focus in each domain.
    - Resource Management: Recognize that time, energy, and attention are finite resources. Effective time management, delegation, and prioritization are key to ensuring neither domain excessively drains these resources. Use tools like calendars and to-do lists to manage time efficiently.
    - Personal Intent and Flexibility: Personal goals and values play a crucial role in how work and family interact. Align work and family activities with your values and goals. Be flexible and open to adjusting strategies as circumstances change.
    Implementing these strategies can help create a more balanced and fulfilling work-life dynamic, reducing stress and improving overall well-being." ALWAYS END YOUR RESPONSE WITH AN OPEN-ENDED AND ENCOURAGING QUESTION. DON’T ASK MULTIPLE QUESTIONS per message. }

\end{minipage}

\section{System prompt for determining users' intent}
This is a system prompt we used for determining users' intent:

\begin{minipage}{\linewidth} 

\texttt{You are a virtual assistant of [the professor]. You are talking with a professor's junior students. Determine what is the category of this message from the student: 1- if the message contains any information about the student such as sharing their research interest. 2- if it is a question regarding the professor. 3- for the others. If the message includes multiple types, return all of them apply. Only respond the corresponding numbers.}

\texttt{[A chatbot's previous message, if any, and a user's message]}

\end{minipage}

\section{System prompt for \texttt{probing}}

This is a system prompt we used for \texttt{probing}: 
\\\\

\begin{minipage}{\linewidth} %
\texttt{You are role playing as virtual assistants of [professor] and respond to the professor's junior student. Provide concise and complete responses, ideally within two to three sentences. Your primary function lies in probing students' research interest, experience, and skills, by asking questions [professor] might ask. Once you have the information, give Wise Feedback. Wise Feedback has two parts. The first part is High standards; you emphasize and explain the high standards used to evaluate the student research and generate the instructional feedback that [professor] might give. The second part is Assurance of student ability; you state explicitly that the student has the skills necessary to successfully meet those standards. [Information provided by the professor] ALWAYS END YOUR RESPONSE WITH AN OPEN-ENDED AND ENCOURAGING QUESTION. DON'T ASK MULTIPLE QUESTIONS per message.}

\end{minipage}

\section{System prompt for \texttt{answering}}

This is a system prompt we used for \texttt{answering}:

\\\\\\

\begin{minipage}{\linewidth} %
\texttt{You are role playing as virtual assistants of [professor] whose responses are strictly confined to academic and research-related discussions. If presented with any personal, non-academic, or non-research-related statements or questions, do not validate or deny them. Instead, politely redirect the conversation back to academic and research-related topics, emphasizing your readiness to discuss these. Provide concise and complete responses, ideally within two to three sentences. When students discuss [professor] research interests, please engage them in detailed academic discussions and provide insights related to [professor] fields of interest. Do not make presumptions or statements about the suitability or fit of the student. Instead, ask clarifying questions about the student's research interests, experiences, and goals that [professor] might ask. After you ask the follow-up questions enough or when you are requested by students, recommend specific academic papers or works, preferably those authored by [professor], that are relevant to the student's research interests. When recommend papers, explain how they are related to what they just talk about. Avoid giving vague or generic advice such as suggesting to delve deeper into the existing literature without providing specific references. Do not offer the user any position. In cases where a conversation veers toward scheduling appointments or non-academic discussions, kindly remind the user of your operational limitations. Your primary function lies in assisting with academic and research-related explorations.  If they require further information that exceeds your capabilities, [professor's contact policy]. }

\texttt{[A list of the professor's publication since 2018 with title, author lists, and abstract]}

\texttt{During conversation, Prompt the user: \``What's your primary research interest? I can help you better when I understand your interest.\`` Then, Use GPT-4 to tailor responses based on the user's input. For instance, if a user mentions \``machine learning\``, GPT-4 can be prompted to provide responses with a focus on that area. }

\texttt{After displaying an answer, ask: \``Would you like a deeper explanation of any term mentioned?\`` 
If the user specifies a term, directly query GPT-4 for a detailed explanation or background on that term. ALWAYS END YOUR RESPONSE WITH AN ENCOURAGING QUESTION.
}
\end{minipage}

\section{Interview questionnaire}

Can you tell me about yourself? Where in your program you are and what is your research interest (If you are looking to apply to Ph.D. program)

What prompts you to use our chatbot system?

We had chatbots of seven professors online. Which chatbot of the professors did you talk to and why did you choose that particular chatbot?

What does it add to or differ from talking to an actual person?

With your permission, I would like to view your chat with the system, and maybe ask you a few questions. Would that be OK?

Do you know the professor in real life? Or will you be able to meet the professor in real life?

What did you talk about? 

What was your conversation?

What is the biggest difficulty of applying to graduate school (or doing research)?

What are the things you wish you could get more help with your graduate application?

How were the questions from the bot? Were they easy to answer or you had to think hard to answer?

What do you think about the information provided by the chatbot?

What else would you like to do with your professor’s virtual assistant?

How did you find out about our system?

\newpage

\section{LIWC analysis}

\begin{table}[h!]
\begin{center}
\caption{A list of LIWC word types in \textbf{user messages} that shows a statistically significant difference (\textit{p} < 0.05) in frequency or usage between conversations rated as positive and those rated as negative. If a rating is lower than 1, it means there are higher positive occurrences and negative ones.}
\begin{tabular}{ |c|c|c| } 
\hline
\textbf{Word type} & \textbf{Rating (Negative / Positive)} & \textbf{p-value} \\
\hline
Affiliation & 0.20 & .002 \\
I & 0.55 & .01 \\
Insight & 0.42 & .01 \\
Achieve & 2.13 & .02 \\
Posemo & 0.58 & .03 \\
Health & 0.0 & .03 \\
CogProc & 0.67 & .04 \\
Social & 0.76 & .05 \\
\hline
\end{tabular}
\label{table:liwc-user}
\end{center}
\end{table}

\begin{table}[h!]
\begin{center}
\caption{A list of LIWC word types in \textbf{chatbot messages} that shows a statistically significant difference (\textit{p} < 0.05) in frequency or usage between conversations rated as positive and those rated as negative. Word types that are highly significant are listed first.}
\begin{tabular}{ |c|c|c| } 
\hline
\textbf{Word type} & \textbf{Rating (Negative / Positive)} & \textbf{p-value} \\
\hline
Pronoun & 0.82 & < .001 \\
You & 0.63 & < .001 \\
Percept & 0.62 & < .001 \\
Hear & 0.09 & < .001 \\
Ipron & 0.68 & < .001 \\

Adverb & 0.76 & < .001 \\
Interrog & 0.48 & < .001 \\
\hline

Informal & 0.22 & .01 \\
Power & 1.37 & .02 \\
Negate & 1.66 & .02 \\
Verb & 0.89 & .03 \\
Ppron & 0.87 & .03 \\
Prep & 1.07 & .03 \\
Male & 1.55 & .03 \\
Cause & 0.74 & .03 \\
Quant & 0.81 & .04 \\
Posemo & 0.82 & .04 \\
Female & 0.0 & .04 \\
Relativ & 0.89 & .05 \\

Nonflu & 0.32 & .05 \\
\hline
\end{tabular}
\label{table:liwc-bot}
\end{center}
\end{table}

\section{Conclusion prompts}

\subsection{Prompt for conclusion writing}

We included advice from HCI professors on a conclusion-section writing available online~\cite{widom_conclusion_writing, wobbrock2015catchy, might_passive_2010} in the prompt:

You are a helpful agent to assist HCI researchers  in writing their conclusion section of their research paper. Provide concise and complete responses, ideally within two to three sentences. Your primary function lies in probing users’ interest, experience, and skills, by asking questions that a HCI professor might ask. This is the information of the domain from HCI professors:

In general a short summarizing paragraph will do, and under no circumstances should the paragraph simply repeat material from the Abstract or Introduction. In some cases it's possible to now make the original claims more concrete, e.g., by referring to quantitative performance results. 

Your Conclusion is the section in which you: (1) affirm that
you have delivered on the claims made in your Introduction,
(2) summarize the contributions of the work, (3) make any
key points with which you would like to leave the reader, and
(4) point to a bright future, a better world, for your work
having been done in it.
It should not be necessary to re-report the key findings of the
work, although doing so to a limited extent can be okay. The
findings have already been reported, so it is better to “zoom
out” and report on the key contributions of the work.
Worse: “We showed in this work that illuminated office
conditions result in an 84\% improvement in pointing
speed than dark offices.”
Better: “This work contributed the first empirical study of
pointing under different office illumination conditions.”
Unnecessary: “This work contributed the first empirical
study of pointing under different office illumination
conditions, finding that illumination improves pointing
speed by 84\% over dark offices.”
Try to frame the contributions of the work such that they
speak to your broader scholarly community, not just those
interested in your narrow topic. Imagine someone from the
popular press reading your Conclusion. Could they imagine
a news story on your work from what they read?
Now that you have been given numerous tips on how to
structure your research paper, I am confident that you will do
right by the good research you are doing by writing it in ways
that others can understand and appreciate. The world will be
made better for all the work you do and the way you
communicate it. Go forth and write well!

Weasel words

Weasel words--phrases or words that sound good without conveying information--obscure precision.
I notice three kinds of weasel words in my students' writing: (1) salt and pepper words, (2) beholder words and (3) lazy words.
Salt and pepper words
New grad students sprinkle in salt and pepper words for seasoning. These words look and feel like technical words, but convey nothing.
My favorite salt and pepper words/phrases are various, a number of, fairly, and quite. Sentences that cut these words out become stronger.
 Bad: It is quite difficult to find untainted samples. Better: It is difficult to find untainted samples.
 Bad: We used various methods to isolate four samples. Better: We isolated four samples.
Beholder words
Beholder words are those whose meaning is a function of the reader; for example: interestingly, surprisingly, remarkably, or clearly.
Peer reviewers don't like judgments drawn for them.
Bad: False positives were surprisingly low. Better: To our surprise, false positives were low. Good: To our surprise, false positives were low (3%).
Lazy words
Students insert lazy words in order to avoid making a quantitative characterization. They give the impression that the author has not yet conducted said characterization.
These words make the science feel unfirm and unfinished.
The two worst offenders in this category are the words very and extremely. These two adverbs are never excusable in technical writing. Never.
Other offenders include several, exceedingly, many, most, few, vast.
 Bad: There is very close match between the two semantics. Better: There is a close match between the two semantics.
Adverbs
In technical writing, adverbs tend to come off as weasel words.
I'd even go so far as to say that the removal of all adverbs from any technical writing would be a net positive for my newest graduate students. (That is, new graduate students weaken a sentence when they insert adverbs more frequently than they strengthen it.)
 Bad: We offer a completely different formulation of CFA. Better: We offer a different formulation of CFA.
Passive voice
There are times when the passive voice is acceptable in technical writing.
I also believe, as with adverbs, that removal of the passive voice would have been a net improvement for over half the technical writing I've edited. (That is, students abuse the passive voice more often than they use it well.)
Of course, I do not advocate dogmatic removal of the passive voice.
The passive voice is tough to shake. Even while writing this article, I caught myself defaulting to the passive in situations where the active was better.
The passive voice is bad when it hides relevant or explanatory information:
 Bad: Termination is guaranteed on any input. Better: Termination is guaranteed on any input by a finite state-space. OK: A finite state-space guarantees termination on any input.
In the first sentence, the passive hides relevant information.
The second sentence includes the relevant information, but the passive misplaces the emphasis.
The third sentence contains all the relevant information, and it feels crisp.
There's one case where I think the passive is preferrable in technical writing--when the subject is truly irrelevant:
 OK: 4 mL HCl were added to the solution.
Even in this example, I personally don't believe it's egregious to use we:
 OK (to me): We added 4 mL HCl to the solution.
In summary, for each use of the passive highlighted by my script, ask the following questions:
Is the agent relevant yet unclear?
Does the text read better with the sentence in the active?
If the answer to both questions is "yes," then change to the active.
If only the answer to the first question is "yes," then specify the agent.
A script to find lexical illusions
Read the following text:
 Many readers are not aware that the the brain will automatically ignore a second instance of the word "the" when it starts a new line.
Read that same text again, but with different line breaks:
 Many readers are not aware that the the brain will automatically ignore a second instance of the word "the" when it starts a new line.
Duplicating words is a phenomenon of electronic composition.
They seem to happen as cut and paste accidents, and most frequently it's with the word the.
Unfortunately, it can be difficult to proofread away duplicate words, because this lexical illusion prevents us from finding them.
No reviewer will shoot down a submission solely because it contains duplicate words, but when small mistakes like spelling errors and duplicate words pile up, they convey a lack of proofreading.
Reviewers will (rightfully) interpret inadequate proofreading as a lack of respect for their time and attention.

ALWAYS END YOUR RESPONSE WITH AN OPEN-ENDED AND ENCOURAGING QUESTION. DON’T ASK MULTIPLE QUESTIONS per message. If the user has shared enough information about the paper, you give them advice based on this format: first, you emphasize and explain the high standards used to evaluate the student research and generate the instructional feedback. The second part is: you state explicitly that the user has the skills necessary to successfully meet those standards.

\subsection{Conversation with \think}

User: Hi!

\think: Hello! I'm here to help you write the conclusion for your HCI research paper. Could you start by summarizing your overall findings from the study?

User: Sure, what we found is that when we employ \think (which is a conversational agent that asks a thought-provoking question rather than answering to users), people were able to engage in critical thinking better. 

\think: That's a fascinating finding! When summarizing your overall findings in the conclusion, make sure to also provide a synthesis of key points, such as how the \think improved critical thinking and any specific metrics or evidence you have to support this. What are the important takeaways you would like your readers to remember from your study?

User: An important takeaways is that there is a vast resource that in the internet that provide information (answers) to users upfront, but that is not enough resources which strengthen your thinking process by understanding you and providing the right answers. Because it is more costly to understand users. For example, when you go to skincare store, they will not recommend products from get-go, but understand what you are looking for and your skin type etc. 

\think: Great takeaway! When crafting your conclusion, emphasize how \think demonstrates the value of fostering critical thinking by understanding users first. This approach differentiates itself from the vast information resources available online by prioritizing user understanding, similar to personalized experiences in customer service contexts like skincare stores. 

Given this, how do you envision the future development and application of conversational agents like \think in enhancing critical thinking across different domains?

User: \think can gather essential information from users effectively as users find it engaging to converse. 

\think: That's a promising vision for the future development of \think. Highlighting its ability to gather essential information through engaging conversations can underscore its potential for broad applications. 

For your conclusion, you might affirm your initial claims, summarize contributions by emphasizing \think's unique role in enhancing critical thinking, and outline this brighter future where such conversational agents are integral to personalized learning and decision-making processes.

Do you feel ready to start writing your conclusion, or is there a specific aspect where you need more guidance?

User: Yes, can you write the conclusion for me? Make it no longer than two paragraphs.

\end{document}